\documentclass{ieeeaccess}

\usepackage{float}
\usepackage{cite}
\usepackage{amsmath,amssymb,amsfonts}
\usepackage{float}
\usepackage{graphicx}
\usepackage{textcomp}
\usepackage{bm}
\usepackage{array}  
\usepackage{multirow}  

\usepackage{array}    
\usepackage{calc}  

\makeatletter

\AtBeginDocument{\DeclareMathVersion{bold}
\SetSymbolFont{operators}{bold}{T1}{times}{b}{n}
\SetSymbolFont{NewLetters}{bold}{T1}{times}{b}{it}
\SetMathAlphabet{\mathrm}{bold}{T1}{times}{b}{n}
\SetMathAlphabet{\mathit}{bold}{T1}{times}{b}{it}
\SetMathAlphabet{\mathbf}{bold}{T1}{times}{b}{n}
\SetMathAlphabet{\mathtt}{bold}{OT1}{pcr}{b}{n}
\SetSymbolFont{symbols}{bold}{OMS}{cmsy}{b}{n}
\renewcommand\boldmath{\@nomath\boldmath\mathversion{bold}}}

\makeatother

\def\BibTeX{{\rm B\kern-.05em{\sc i\kern-.025em b}\kern-.08em
    T\kern-.1667em\lower.7ex\hbox{E}\kern-.125emX}}

\begin{document}
\history{Date of publication xxxx 00, 0000, date of current version xxxx 00, 0000.}
\doi{10.1109/ACCESS.2024.0220000}

\title{Breast Cancer Classification with Enhanced Interpretability: DALAResNet50 and DT Grad-CAM}
\author{\uppercase{Suxing Liu}\authorrefmark{1,2},
\uppercase{GALIB MUHAMMAD SHAHRIAR HIMEL}\authorrefmark{3}, \uppercase{Jiahao Wang}
\authorrefmark{3} } 

\address[1]{Department of IT Engineering, Mokwon University, Daejeon 35349, South Korea}

\address[2]{School of Digital Arts,Jiangxi Arts \&Ceramics Technology Institute, Jindezhen 33001, China }

\address[3]{School of Computer Sciences, Universiti Sains Malaysia, Penang 11800, Malaysia}

\markboth
{Author \headeretal: Preparation of Papers for IEEE TRANSACTIONS and JOURNALS}
{Author \headeretal: Preparation of Papers for IEEE TRANSACTIONS and JOURNALS}

\corresp{Corresponding author: Suxing Liu (bentondoucet@gmail.com)}

\begin{abstract}

Automatic classification of breast cancer in histopathology images is crucial for accurate diagnosis and effective treatment planning. Recently, classification methods based on the ResNet architecture have gained prominence due to their ability to improve accuracy significantly. This is achieved by employing skip connections to mitigate vanishing gradient issues, enabling the integration of low-level and high-level feature information. However, the conventional ResNet architecture faces challenges such as data imbalance and limited interpretability, which necessitate cross-domain knowledge and collaboration among medical experts. To address these challenges, this study proposes a novel method for breast cancer classification: the Dual-Activated Lightweight Attention ResNet50 (DALAResNet50) model. This model integrates a pre-trained ResNet50 architecture with a lightweight attention mechanism, embedding an attention module in the fourth layer of ResNet50, and incorporates two fully connected layers with LeakyReLU and ReLU activation functions to enhance feature learning capabilities. Extensive experiments conducted on the BreakHis, BACH, and Mini-DDSM datasets demonstrate that DALAResNet50 outperforms state-of-the-art models in accuracy, F1 score, IBA, and GMean, particularly excelling in classification tasks involving imbalanced datasets. Furthermore, the proposed Dynamic Threshold Grad-CAM (DT Grad-CAM) method provides clearer and more focused visualizations, enhancing interpretability and assisting medical experts in identifying key features.
\end{abstract}


\titlepgskip=-21pt

\maketitle

\section{Introduction}
\label{sec:introduction}

Breast cancer has become a leading cause of mortality, particularly among women, with its incidence rising continuously, raising global health concerns \cite{b1}. Accurate diagnosis and timely treatment are pivotal in decreasing the mortality rate associated with breast cancer. Yet, medical practices face challenges such as insufficient healthcare resources and human error in interpretation, which impede the timely and precise detection and diagnosis of breast cancer. Against this backdrop, intelligent diagnostic assistance technologies have emerged as innovative solutions in the battle against breast cancer. Recent advances in deep learning have revolutionized breast cancer diagnosis, overcoming the limitations of traditional methods such as radiomics \cite{Prinzi2024} and \cite{Siviengphanom2023}. Conventional methods often rely on separate steps for feature extraction, selection, and classification, as illustrated by multivariate radiomic time series analysis in dynamic contrast-enhanced magnetic resonance imaging (DCE-MRI) sequences \cite{Prinzi2024}  and the use of global radiomic features from mammography to predict difficult-to-interpret normal cases \cite{Siviengphanom2023}. Deep learning automates feature extraction and classification through neural networks, reducing redundancy and enhancing efficiency, thus significantly improving diagnostic accuracy and workflow \cite{b3,b4,b5,b6,b7}. A pivotal innovation in this domain is developing and applying Convolutional Neural Networks (CNNs), specialized deep learning models adept at processing data, such as medical images, with a grid-like topology. In 1998, LeCun et al. \cite{b8} introduced Convolutional Neural Networks (CNNs), which progressively extract abstract features from images through multiple levels of convolutional and pooling operations. This process enables CNNs to achieve classification based on these extracted features. CNNs comprise convolutional, pooling, and fully connected layers, collaboratively facilitating the robust classification of medical images. The convolutional layer identifies local image features; the pooling layer reduces data dimensions while retaining critical information; and the fully connected layer maps these features to output classes. Despite deep learning's significant achievements in medical image classification, conventional CNN architectures often struggle with imbalanced medical image data.

This study introduces the Dual-Activated Lightweight Attention ResNet50 \cite{b9} (DALAResNet50) model, developed to address the challenge of imbalanced data in CNN architectures. By integrating this module into the fourth residual block of ResNet50, the model enhances attention to essential feature regions while reducing the high computational complexity associated with traditional attention mechanisms,such as SE and CBAM modules, making it suitable for real-time medical image analysis. The DALAResNet50 incorporates three key innovations: adaptive average pooling, which replaces traditional pooling layers to dynamically adjust feature extraction based on input image size, reducing parameter count and improving global feature capture; a dual activation strategy that combines LeakyReLU and ReLU to mitigate the vanishing gradient problem and enhance non-linear feature expression, promoting stable training and effective learning of complex patterns; and element-wise multiplication, enabling dynamic feature map adjustment to improve adaptability when handling imbalanced datasets and ensuring critical features in minority classes are retained. These innovations collectively enhance the model’s feature extraction capabilities while significantly reducing computational demands. Additionally, the study introduced Dynamic Threshold Grad-CAM (DT Grad-CAM) for improved interpretability. This innovative approach efficiently manages uneven category distributions and enhances the detection of malignant breast cancer in medical images. The primary contributions of this work are summarized as follows:

(a) We introduce DALAResNet50, a novel breast cancer classification framework designed to address the challenges of data imbalance, limited adaptability, and overfitting commonly associated with the traditional ResNet architecture.

(b) We propose DT Grad-CAM, a refined visualization technique that overcomes the issues of blurriness and lack of prominence found in conventional Grad-CAM methods. This technique enhances the clarity and visibility of salient regions, thereby improving model interpretability and offering deeper insights into the predictive features emphasized by the network.

(c)Extensive experiments conducted on the BrekHis, BACH, and Mini-DDSM datasets demonstrate that DALAResNet50 consistently outperforms eight baseline models across key metrics, including accuracy, F1 score, IBA, and GMean. Our approach exhibits superior performance in classification tasks, particularly on imbalanced datasets. Additionally, DT Grad-CAM visualizations provide more distinct and concentrated highlight regions, further substantiating the robustness of our method.

Organization of the paper:

Section II reviews the latest advancements in applying deep learning for breast cancer diagnosis and Grad-CAM-based Visualization Techniques in Histopathology Image Analysis.

Section III  proposed the DALAResNet50 model approach for classifying breast cancer histopathology images.

Section IV outlines the dataset and implementation specifics of the experimental setup and compares the classification outcomes with those of other leading networks.

Section V  compares the results of Grad-CAM with other CAM methods, highlighting the significant advantages of DT Grad-CAM in visualization and diagnostic accuracy.

Section VI  provides a summary and conclusion
\section{Related works}

This section outlines the relevant work in the field of cancer tissue histopathology image classification, which can be primarily categorized into two areas: Deep Learning-Based Breast Cancer Image Classification and Grad-CAM-based Visualization Techniques in Histopathology Image Analysis.
\subsection{Deep Learning-Based Breast Cancer Image Classification}
Convolutional Neural Networks (CNNs) are widely used in image processing, excelling in tasks such as breast cancer histopathology image classification. When trained on large datasets, models like AlexNet, Inception, VGGNet, DenseNet, and ResNet have demonstrated substantial capabilities. AlexNet\cite{b16}, marked a significant milestone, showcasing CNN’s potent capabilities. Modified versions, like the one by Omonigho et al. \cite{b17} on the MIAS mammography database, achieved notable accuracy, outperforming traditional methods. Inception, developed by Google, focuses on computational and parameter efficiency. Nazir et al.’s hybrid CNN-Inception-V4 model \cite{b18} showed high accuracy and sensitivity, despite some specificity and computational efficiency limitations. VGGNet excels in image classification but demands intensive computational resources \cite{b19}. DenseNet \cite{b20}, known for efficient gradient flow through dense connectivity, was leveraged by Jiménez Gaona et al. \cite{b21} for breast cancer detection, reducing parameters, and improving performance. ResNet, introduced by He et al. in 2015 \cite{b9}, tackles the vanishing gradient problem in deep networks, with Khikani et al. \cite{b22} enhancing it using Res2Net blocks for improved breast cancer classification.

To further improve the effectiveness of these models, transfer learning involves applying pre-trained models to new tasks, and optimizing them for specific needs like breast cancer detection \cite{b23,b24,b25,b26,b27,b28,b29}.  Ismail et al. \cite{b30} confirmed that transfer learning, using models like VGG16 and ResNet50, quickens convergence and outperforms training from scratch. Muhammad et al. \cite{b31} used pre-trained models such as ResNet50, ResNet101, VGG16, and VGG19, finding that ResNet50 performed best with an accuracy rate of 90.2\%, an AUC of 90.0\%, and a recall rate of 94.7\%. By optimizing ResNet50 in transfer learning, researchers aim to address challenges with imbalanced datasets, reducing biases towards categories with more instances and overfitting in minority classes.

Building on the strengths of these models, the attention mechanism \cite{b32,b33,b34} in deep learning allows them to focus on important regions within medical images, enhancing accuracy and efficiency. Attention mechanisms help distinguish critical features in breast cancer classification, improving differentiation between malignant and benign tissues. Researchers have developed various systems incorporating attention mechanisms. For instance, Vardhan et al. \cite{b35} utilized CNNs with interpretable AI and attention mechanisms for enhanced detection and classification. Xu et al. \cite{b36}designed a system using a modified DenseNet with attention mechanisms, evaluated on the BreakHis dataset. Wang et al. \cite{b37} proposed a model combining deep transfer learning and visual attention mechanisms, achieving high-precision results surpassing other state-of-the-art algorithms.

Despite the benefits, attention mechanisms face challenges such as limitations in global feature extraction, insufficient feature priority precision, inefficiency of activation functions, and lack of feature adaptability. This study designed a lightweight attention module incorporating adaptive average pooling, a two-layer fully connected network for channel recalibration, dual activation functions (LeakyReLU and ReLU), and element-wise multiplication. These innovations aim to enhance model performance while reducing computational resource requirements.

\subsection{Grad-CAM-based Visualization Techniques in Histopathology Image Analysis}

Gradient-weighted Class Activation Mapping (Grad-CAM)  is a widely used technique for visualizing an image's regions most relevant to a neural network’s predictions. This technique has been extensively explored in histopathology image analysis to improve model interpretability and provide insights into the decision-making process of deep learning models \cite{b59}.

In histopathology, accurate interpretation of tissue images is crucial for diagnosing diseases such as cancer. Traditional Grad-CAM methods generate heatmaps by computing the gradients of the target class concerning the feature maps of the last convolutional layer, highlighting the regions of the image that contribute most to the classification decision. This allows pathologists to understand which areas of the tissue image the model focuses on, aiding in verifying the model’s predictions.

However, conventional Grad-CAM has limitations, particularly in dealing with noisy activations and less distinct regions, which can hinder the clarity and precision of the visualizations. Various enhancements have been proposed to address these challenges. For instance, Guided Grad-CAM combines Grad-CAM with guided backpropagation to produce high-resolution class-discriminative visualizations, offering finer details of the important regions \cite{b60}. Additionally, Score-CAM removes the dependency on gradients by using the target class score, improving the heatmaps' stability and robustness\cite{b61}.    ``

To further enhance the interpretability and effectiveness of visual explanations in histopathology image analysis, we implemented the Dynamic Threshold Grad-CAM (DT Grad-CAM) method. DT Grad-CAM addresses the limitations of conventional Grad-CAM by applying adaptive thresholding using Otsu’s method, which dynamically determines the optimal threshold for highlighting significant regions. This approach reduces noise and enhances the clarity of the visualizations, making it easier to identify and interpret the critical features that contribute to the model's predictions. By focusing on the most relevant areas and minimizing the influence of less important regions, DT Grad-CAM provides more precise and interpretable heatmaps, which are particularly beneficial in clinical settings where accurate and transparent decision-making is essential.

\begin{table}[t]
    \centering
    \caption{The ResNet50 Model}
    \label{mnt}
    \begin{tabular}{|c|c|c|}
    \hline
    Module Name & Output Size & 50 Layers \\
    \hline
    Conv1 & $112\times112$ & $7\times7$, 64 stride=2 \\ \hline
    Conv2\_x & $56\times56$ & 
    $\begin{Bmatrix}
        1\times1, & 64  \\
         3\times3 & 64  \\
        1\times1 & 256 
    \end{Bmatrix}$ x3 \\ \hline
    Conv3\_x & $28\times28$ & 
    $\begin{Bmatrix}
        1\times1, & 128  \\
         3\times3 & 128  \\
        1\times1 & 512 
    \end{Bmatrix}$ x4 \\  \hline
    Conv4\_x & $14\times14$ & 
    $\begin{Bmatrix}
        1\times1, & 256  \\
         3\times3 & 256  \\
        1\times1 & 1024 
    \end{Bmatrix}$ x6 \\ 
    \hline
    Conv5\_x & $7\times7$ & 
    $\begin{Bmatrix}
        1\times1, & 512  \\
         3\times3 & 512  \\
        1\times1 & 2048 
    \end{Bmatrix}$ x3 \\ \hline
   Pooling & $1\times1$ & 
    Average Pool, 1000-d FC.
    \\ 
    \hline
      \multicolumn{3}{|p{8.0cm}|}{Note: The term "1000-d FC" in the pooling layer refers to a fully connected layer consisting of 1000 neurons.} \\ 
    \hline
    \end{tabular}
\end{table}

\begin{figure*}
    \centering
    \includegraphics[width=\textwidth]{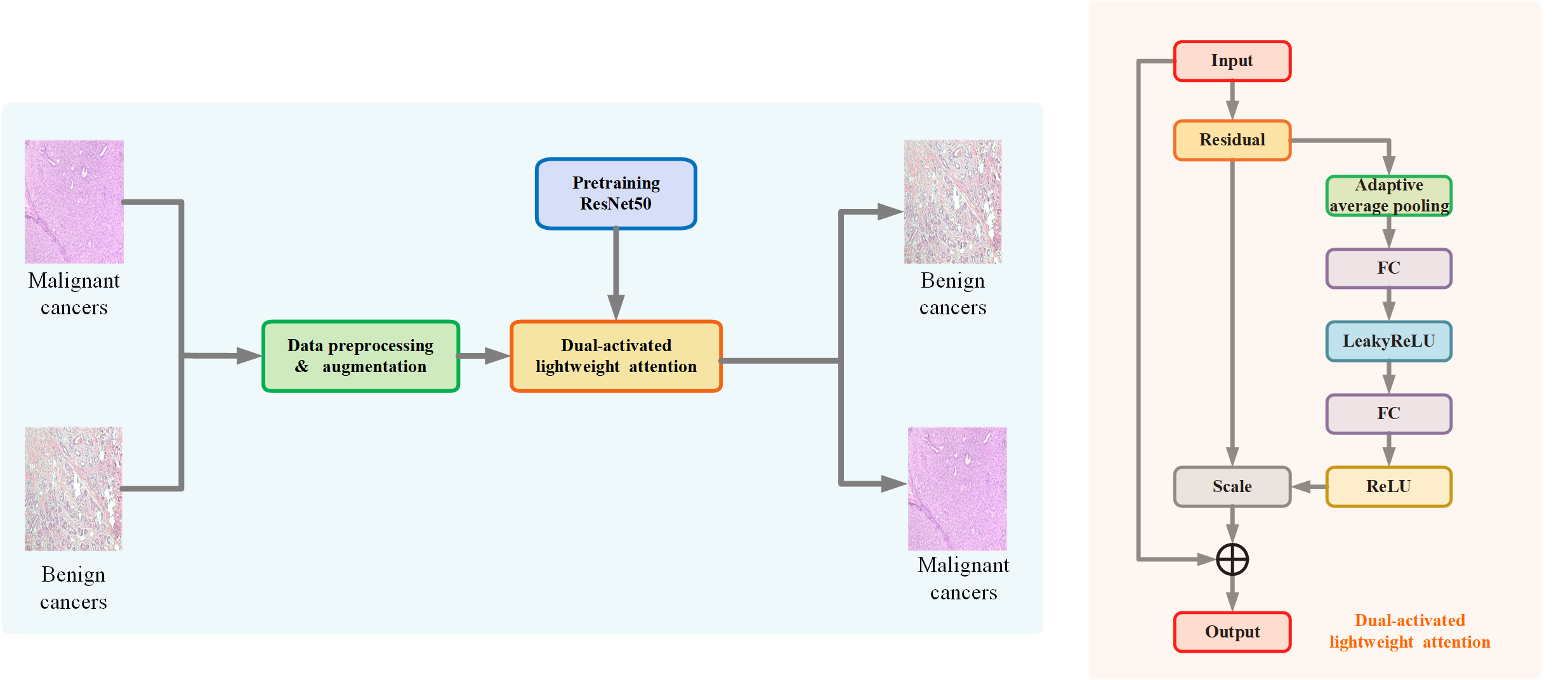} 
    
    \renewcommand{\figurename}{Fig.}
    \caption{The framework of DALAResNet50 model}
    \label{figc}
\end{figure*}

\begin{figure*}[t]
    \centering
    \includegraphics[width=0.7\textwidth,keepaspectratio]{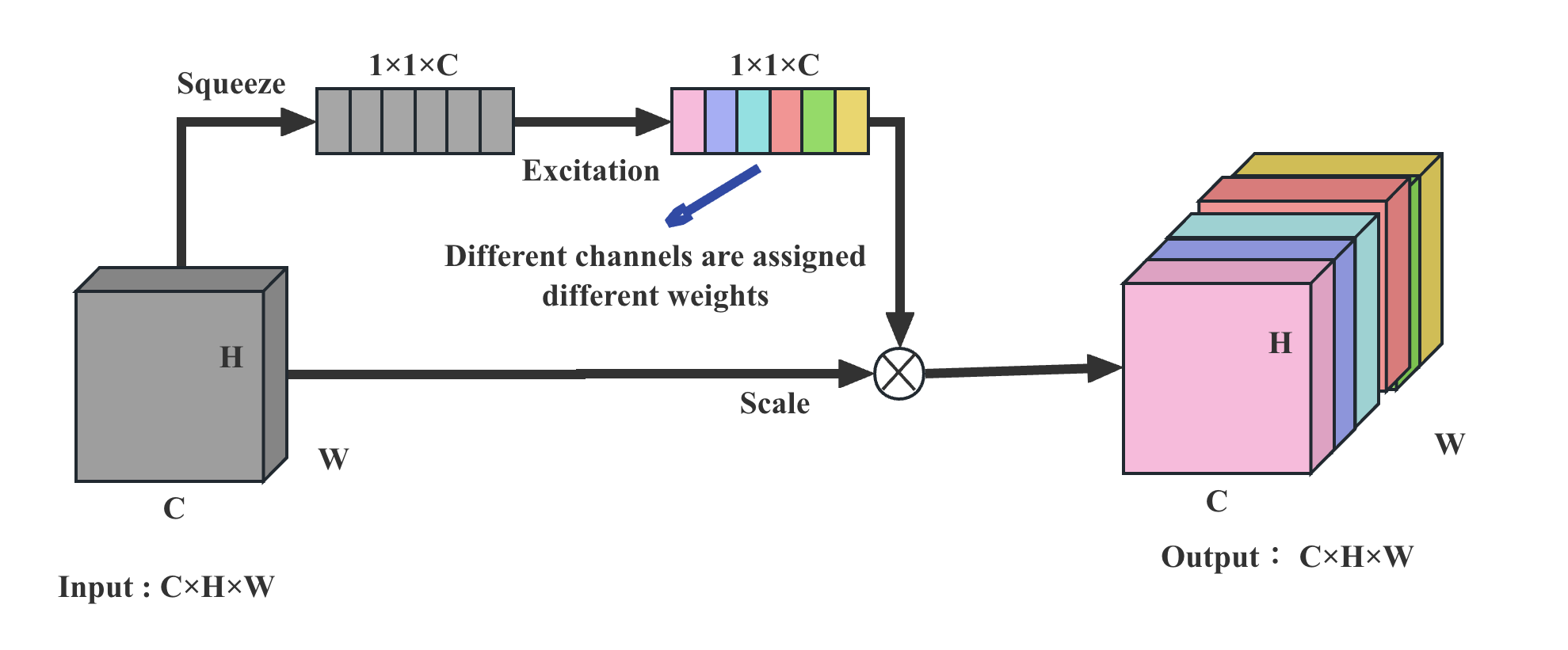}
    \renewcommand{\figurename}{Fig.}
    \caption{Visualization of DALAResNet50 model}
    \label{figA}
\end{figure*}

\section{Methods}
\subsection{ ResNet50 }

The ResNet50 model, as detailed in \cite{b9}, features a relatively complex architecture with a 50-layer deep network structure, as shown in Table 1. The main module consists of numerous residual blocks with multiple convolutional layers and unique identity mappings. Notably, the ResNet50 model introduces the concept of residual learning, which addresses the vanishing gradient problem in deep networks,  enabling the construction of even deeper neural networks. This design breakthrough allows for promising performance. Consequently, this paper adopts a strategy based on the pre-trained ResNet50 model, considering the efficient utilization of data and computational resources in model construction. The benefits of pre-training include leveraging high-level feature representations learned from large-scale datasets for medical image classification tasks. This enables the model to achieve superior classification results rapidly, even with limited data and computational capabilities.

\begin{table*}[t]
\centering
\caption{Performance of the Lightweight Dual-Activation Attention Module Integrated into Different Residual Blocks on the BreakHis Dataset }
\label{your_table_label}
\scriptsize 
\begin{tabular}{cccccccc}
\hline
\textbf{Network Models} & \multicolumn{3}{c}{\textbf{40X}} & \multicolumn{3}{c}{\textbf{100X}} \\ 
\cline{2-7}
 & \textbf{Accuracy} & \textbf{F1 Score} & \textbf{AUC-ROC} & \textbf{Accuracy} & \textbf{F1 Score} & \textbf{AUC-ROC} \\ 
\hline
layer-1 & 0.982 & 0.964/0.982 & 0.995& 0.964 & 0.929/0.964 & 0.990 \\ 
layer-2 & 0.962 & 0.925/0.962 & 0.993 & 0.966 & 0.933/0.966 & 0.995\\ 
layer-3 & 0.983 & 0.966/0.983 & 0.996 & 0.970 & 0.941/0.970 & 0.994 \\ 
\textbf{layer-4} & \textbf{0.985} & \textbf{0.970/0.985} & \textbf{0.998} & \textbf{0.987} & \textbf{0.974/0.987} & \textbf{0.997} \\ 
\hline
\end{tabular}

\vspace{0.3cm} 

\begin{tabular}
{cccccccc}
\hline
\textbf{Network Models} & \multicolumn{3}{c}{\textbf{200X}} & \multicolumn{3}{c}{\textbf{400X}} \\ 
\cline{2-7}
 & \textbf{Accuracy} & \textbf{F1 Score} & \textbf{AUC-ROC} & \textbf{Accuracy} & \textbf{F1 Score} & \textbf{AUC-ROC} \\ 
\hline

 layer-1 & 0.967 & 0.935/0.967 & 0.994 & 0.923 & 0.852/0.923 & 0.990 \\ 
layer-2 & 0.965 & 0.931/0.965 & 0.994 & 0.935& 0.874/0.935 & 0.991 \\ 
layer-3 & 0.963 & 0.927/0.963 & 0.991 & 0.933 & 0.870/0.933 & 0.989 \\ 
\textbf{layer-4} & \textbf{0.979} & \textbf{0.958/0.979}& \textbf{0.996 }& \textbf{0.943} & \textbf{0.891/0.944} & \textbf{0.999} \\ 
\hline
\end{tabular}
\end{table*}

\subsection{Dual-activated lightweight attention ResNet50 }

We utilize a pre-trained ResNet50 framework and incorporate a custom-designed lightweight dual-activation attention module into its fourth residual block. This module aims to pinpoint small-scale cell nuclei characteristic of breast cancer and extract relevant semantic information, thereby significantly enhancing the model’s ability to learn and recognize features. As shown in Table 2, the fourth residual block integration achieved superior results in terms of Accuracy\cite{b46,b47,b48}, F1\cite{b49,b52}, and AUC-ROC\cite{Almarri2024, Murty2024},  demonstrating the effectiveness of this choice. Table 2 references the BreakHis dataset and related experimental details, which are thoroughly discussed in section IV. During the optimization process of the DALAResNet50 model, the focus is primarily on pooling layers, activation layers, and nonlinear feature adjustments to enhance the model’s performance and adaptability, as illustrated in Fig. 1. Fig. 2 depicts the visualization of the DALAResNet50 approach. The specific optimization steps are as follows:

First, traditional pooling layers are replaced with adaptive average pooling to address spatial variations in input feature maps. This adjustment optimizes the network's parameter structure, enabling the model to process global contextual information more effectively. The adaptive average pooling operation is defined as follows:

\begin{equation}
y_i = \frac{1}{H_i \times W_i} \sum_{h=1}^{H_i} \sum_{w=1}^{W_i} x_{i,h,w}
\end{equation}

where \( y_i \) is the output for the \(i\)-th feature map, \( H_i \) and \( W_i \) are the height and width of the \(i\)-th feature map, respectively, and \( x_{i,h,w} \) represents the value at position \((h,w)\).

Next, the model employs a dual activation function strategy in the activation layers. Initially, the LeakyReLU function is used to combat vanishing gradients and improve training efficiency. The LeakyReLU activation function is defined as follows:

\begin{equation}
f(x) = \begin{cases}
x & \text{if } x > 0 \\
\alpha x & \text{otherwise}
\end{cases}
\end{equation}

where \( f(x) \) represents the LeakyReLU function, and \( \alpha \) is a small constant (typically 0.01) that allows a small gradient when the input \( x \) is negative. Subsequently, the ReLU function is applied to further optimize model performance and enhance feature representation. The ReLU activation function is defined as follows:

\begin{equation}
f(x) = \max(0, x)
\end{equation}

where \( f(x) \) is the ReLU function, which outputs \( x \) if \( x \) is positive and 0 otherwise.

Finally, the introduction of element-wise multiplication operations further enhances the model's feature adaptability, allowing for dynamic, nonlinear adjustments of features based on specific inputs. This step significantly increases the network's flexibility and information utilization, enabling the model to adapt its processing flow according to the particular circumstances of the input features. This approach notably improves the model's feature processing capability, leading to higher accuracy and efficiency in classifying breast cancer images. The element-wise multiplication operation is defined as follows:

\begin{equation}
z_i = x_i \cdot y_i
\end{equation}

where \( z_i \) is the result of the element-wise multiplication of the \(i\)-th feature map, and \( x_i \) and \( y_i \) are the corresponding elements of the feature maps being multiplied.
 This operation allows the model to adjust the features dynamically, enhancing its ability to capture complex patterns and interactions within the data.

\section{Experiment}
\subsection{Implementation Details}

Our implementation leverages PyTorch 2.0.1 paired with Python 3.10 and trains on a single NVIDIA RTX A4000 GPU with 16GB video memory. We set a lower initial learning rate of 0.0001 for the small and imbalanced medical imaging dataset to prevent loss value explosion and ensure proper convergence. The batch size is set to 32, the minimum value considering memory constraints, allowing more frequent updates. The number of iterations is limited to 50, balancing training effectiveness and computational efficiency. A lower dropout rate of 0.25 ensures categories with fewer samples are adequately learned, enhancing the model’s recognition accuracy and generalization.

\subsection{Dataset}

Our network is trained and evaluated on three main datasets: BreakHis \cite{b38}, containing 7,909 breast cancer histopathology images (2,429 benign and 5,429 malignant) at 40×, 100×, 200×, and 400× magnifications, each 700×460 pixels; BACH from ICIAR2018 \cite{Aresta2019}, with 400 HE-stained images in four categories (Normal, Benign, In-situ carcinoma, Invasive carcinoma), each 2048×1536 pixels; and Mini-DDSM \cite{b62}, a reduced version of DDSM with approximately 2,000 annotated mammography images detailing lesion locations and diagnoses. 
 Fig. 3, 4, and 5 depict sample images from the BreakHis, ICIAR2018, and Mini-DDSM datasets. Additionally, Tables 3, 4, and 5 present statistical data for these datasets, including sample sizes and class distributions.

\begin{figure}[t]
    \centering
    \begin{minipage}{0.2\textwidth}
        \centering
        \includegraphics[width=0.8\textwidth, height=0.8\textwidth]{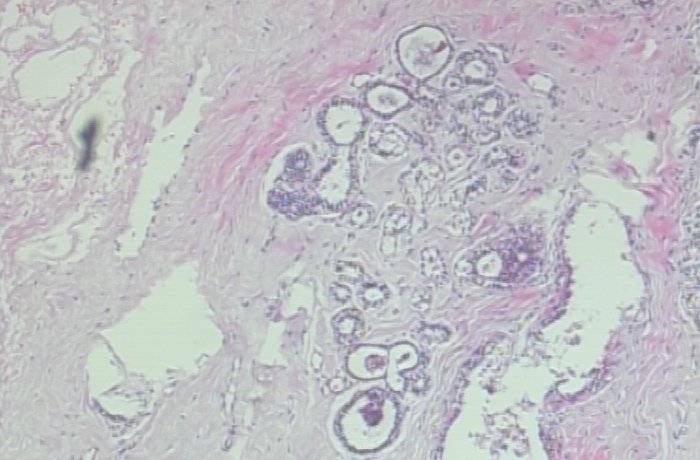}\\
        40x
    \end{minipage}
    \begin{minipage}{0.2\textwidth}
        \centering
        \includegraphics[width=0.8\textwidth, height=0.8\textwidth]{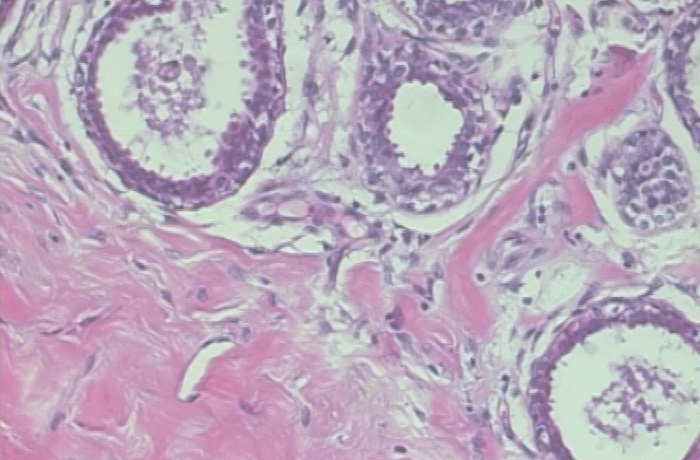}\\
        100x
    \end{minipage}
    
    \vspace{1em}
    
    \begin{minipage}{0.2\textwidth}
        \centering
        \includegraphics[width=0.8\textwidth, height=0.8\textwidth]{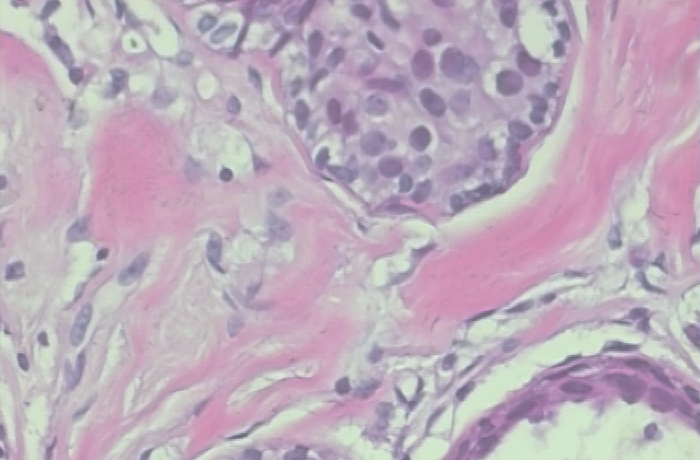}\\
        200x
    \end{minipage}
    \begin{minipage}{0.2\textwidth}
        \centering
        \includegraphics[width=0.8\textwidth, height=0.8\textwidth]{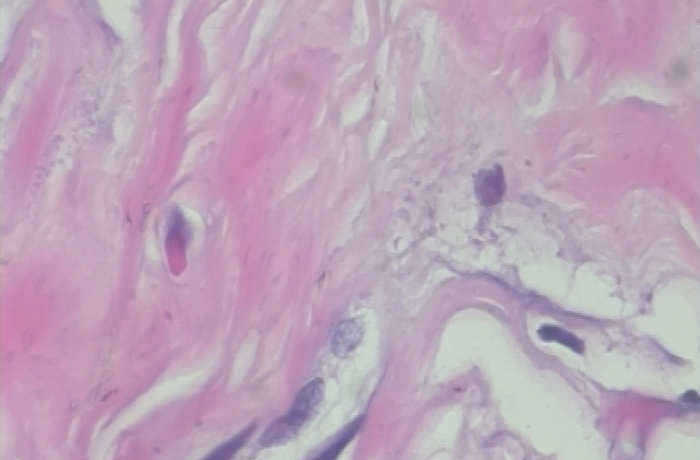}\\
        400x
    \end{minipage}
    
    \renewcommand{\figurename}{Fig.} 
    \caption{Images of Breast Cancer Histopathology from the BreakHis Dataset at Four Magnifications}
    \label{Tal6}
\end{figure}

\begin{figure}[t]
    \centering
    \begin{minipage}{0.2\textwidth}
        \centering
        \includegraphics[width=0.8\textwidth, height=0.8\textwidth]{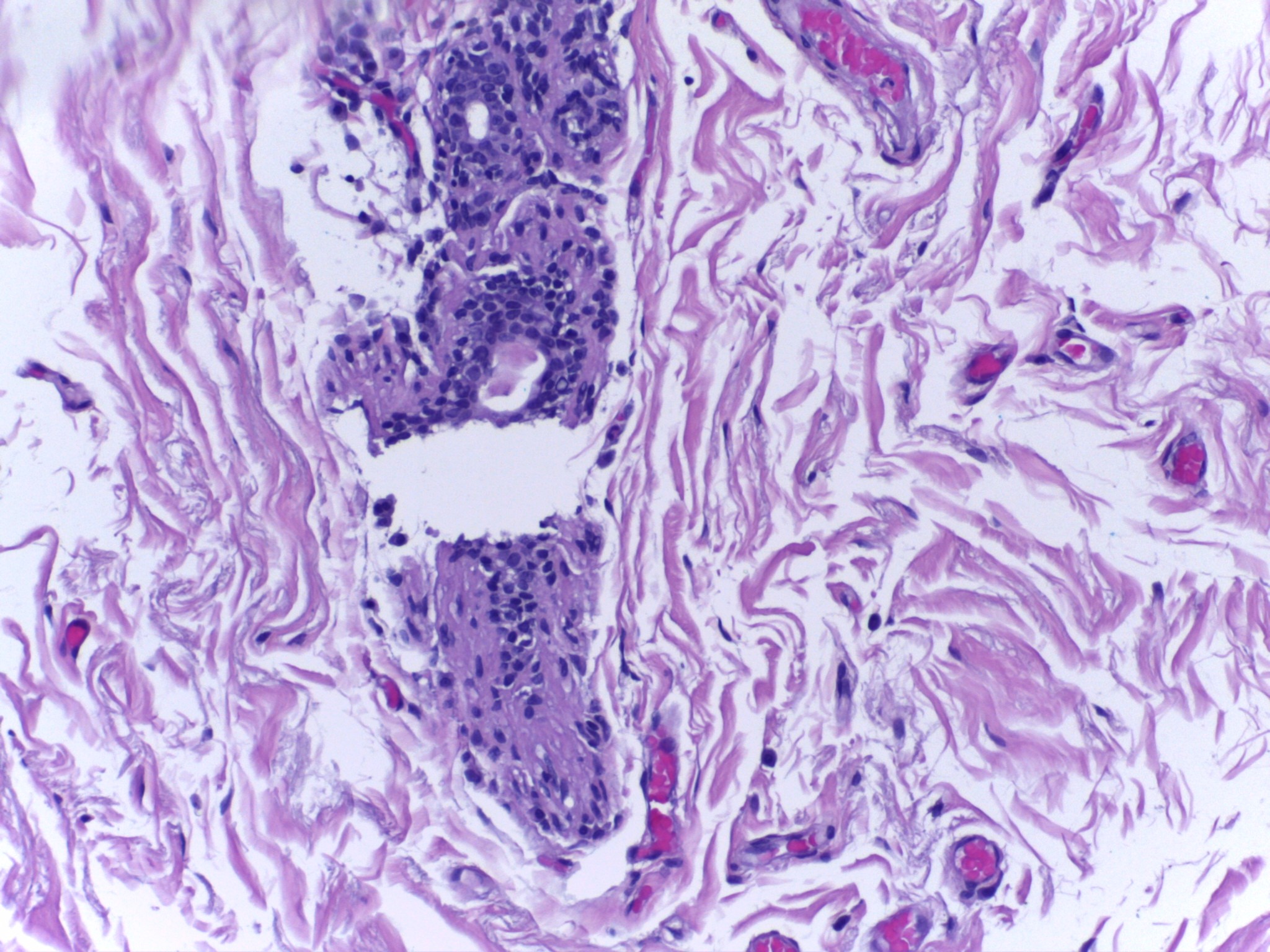}\\
        Normal
    \end{minipage}
    \begin{minipage}{0.2\textwidth}
        \centering
        \includegraphics[width=0.8\textwidth, height=0.8\textwidth]{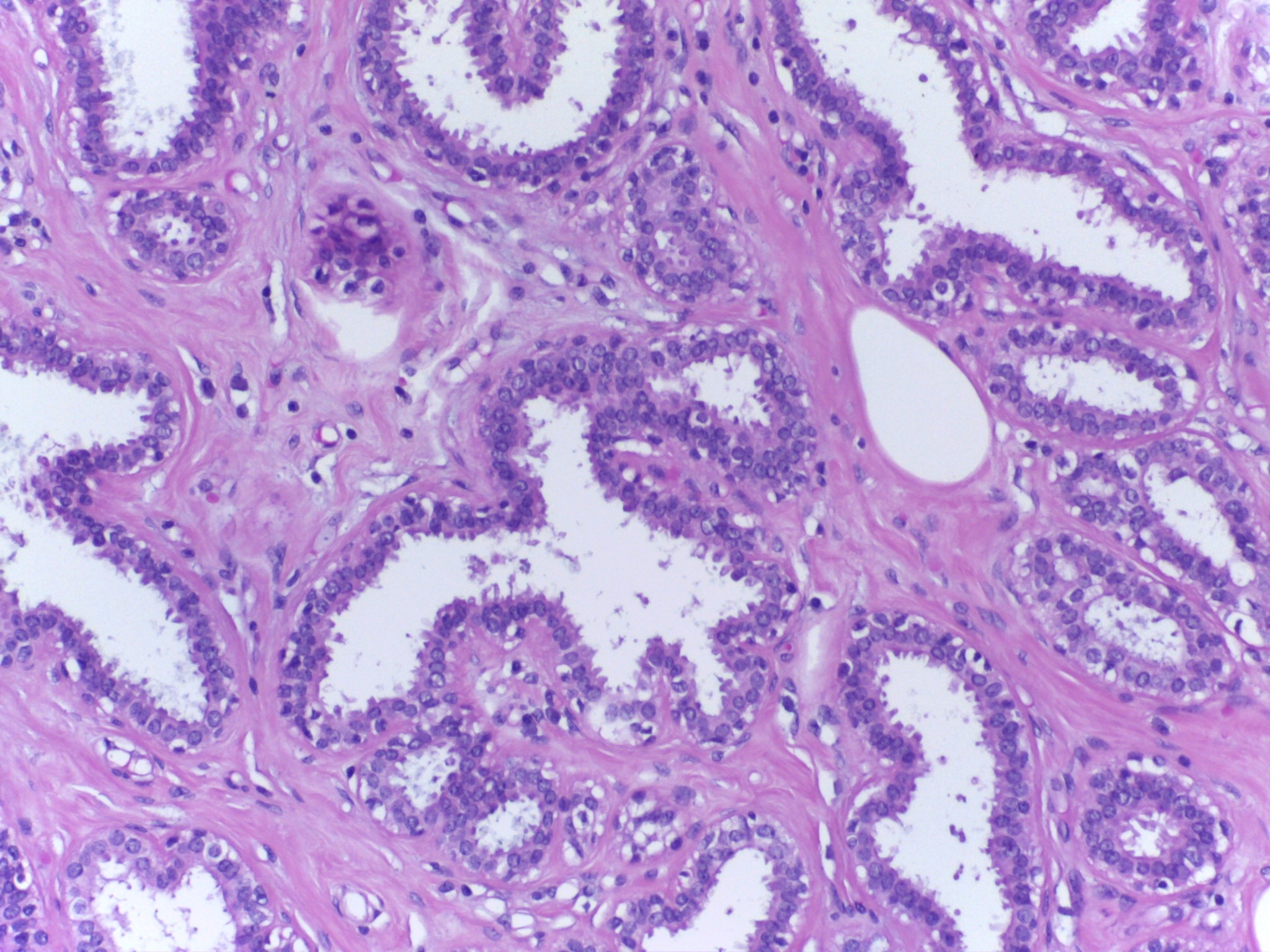}\\
        Benign
    \end{minipage}
    
    \vspace{1em}
    
    \begin{minipage}{0.2\textwidth}
        \centering
        \includegraphics[width=0.8\textwidth, height=0.8\textwidth]{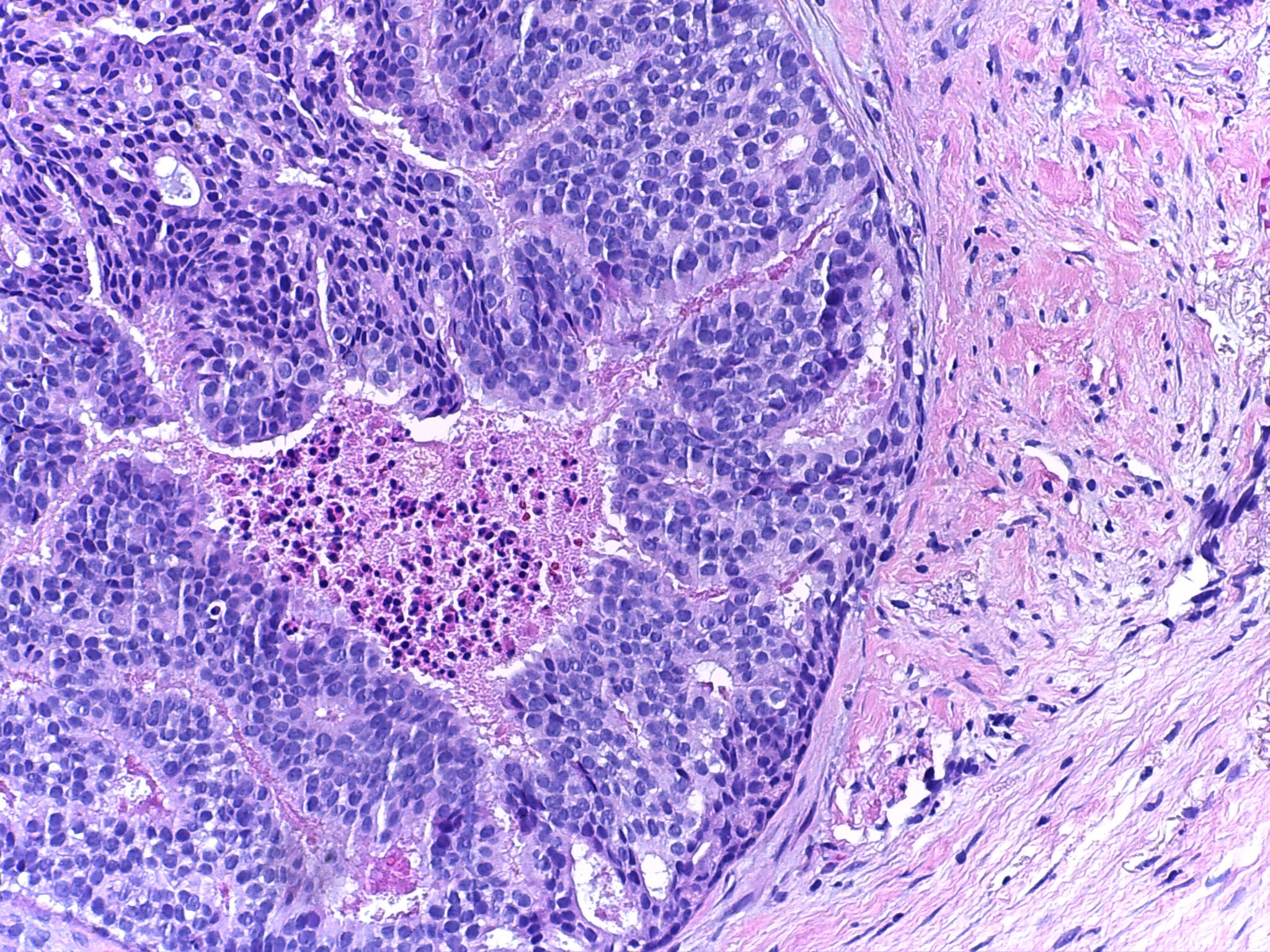}\\
        InSitu
    \end{minipage}
    \begin{minipage}{0.2\textwidth}
        \centering
        \includegraphics[width=0.8\textwidth, height=0.8\textwidth]{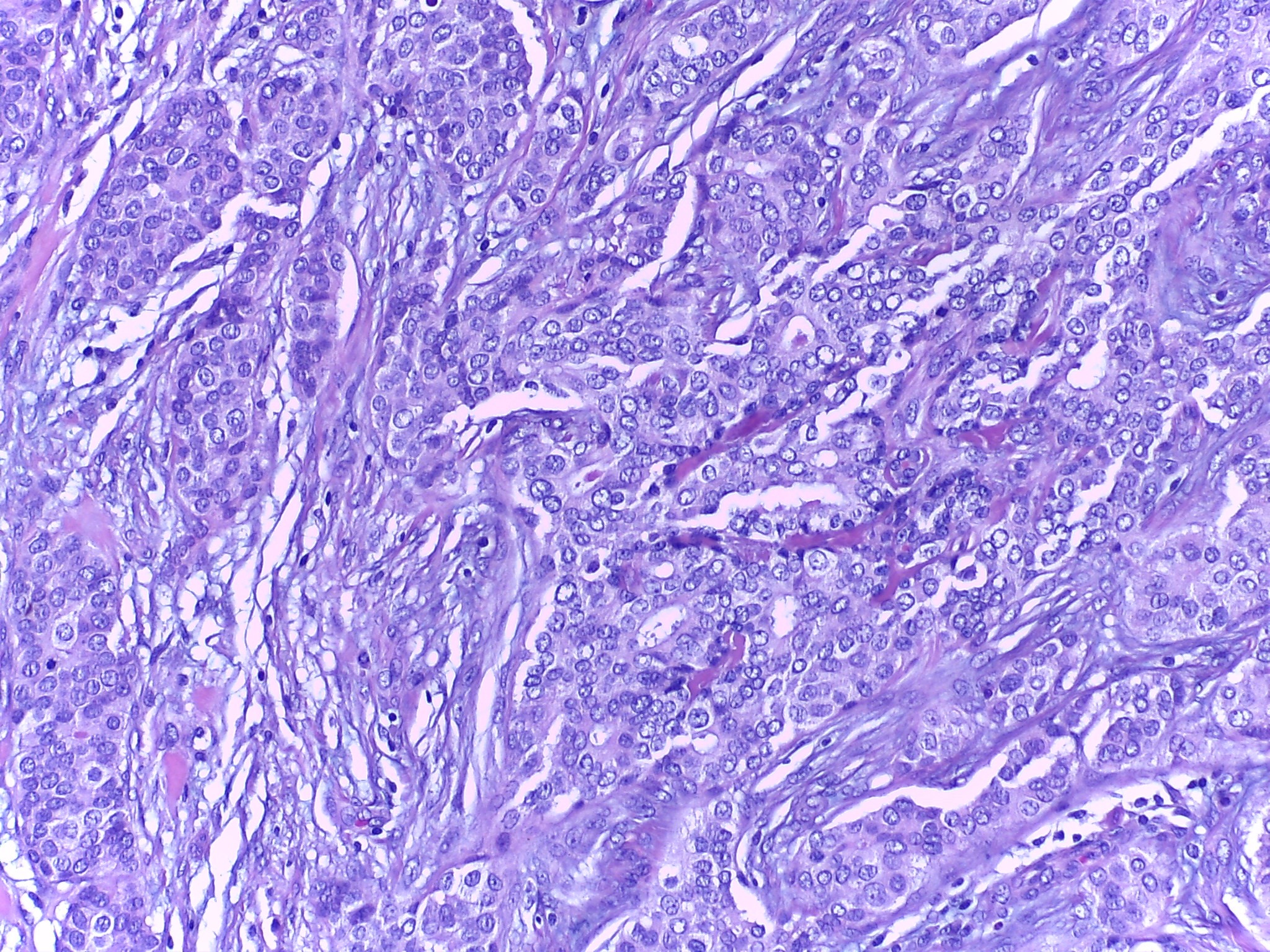}\\
        Invasive
    \end{minipage}
    
    \renewcommand{\figurename}{Fig.} 
    \caption{Images of Breast Cancer Histopathology from the  ICIAR2018  Dataset }
    \label{Tal6}
\end{figure}

\begin{figure}[h]
    \centering
    \begin{minipage}{0.2\textwidth}
        \centering
        \includegraphics[width=0.8\textwidth, height=0.8\textwidth]{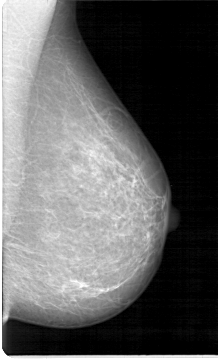}\\
        Normal 
    \end{minipage}
    \begin{minipage}{0.2\textwidth}
        \centering
        \includegraphics[width=0.8\textwidth, height=0.8\textwidth]{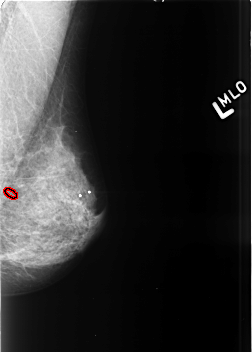}\\
        Benign
    \end{minipage}
    
    \vspace{1em}
    
    \begin{minipage}{0.2\textwidth}
        \centering
        \includegraphics[width=0.8\textwidth, height=0.8\textwidth]{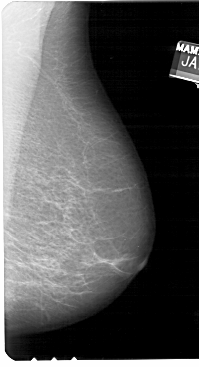}\\
        Malignant
    \end{minipage}
\renewcommand{\figurename}{Fig.} 
    \caption{Images of Breast Cancer Histopathology from the Mini-DDSM Dataset}
    \label{Tal6}
\end{figure}

\begin{table}[H] 
    \caption{BreakHis dataset: benign and malignant image distribution across magnification factors}
    \label{tab:breakhis_distribution}
    \centering
    \begin{tabular}{m{110pt}<{\centering} | cccc}
         \hline
         \multirow{2}{*}{\textbf{Class}} & \multicolumn{4}{c}{Magnification Factor} \\
         \cline{2-5}
         & \textbf{40X} & \textbf{100X} & \textbf{200X} & \textbf{400X} \\
         \hline
         Benign & 625 & 644 & 623 & 588 \\
         Malignant & 1370 & 1437 & 1390 & 1232 \\ 
         Total No of Images & 1995 & 2081 & 2013 & 1820 \\ 
         \hline
    \end{tabular}
\end{table}

\begin{table}[H]
    \centering
    \caption{BACH Dataset Details}
    \label{tab1}
    \begin{tabular}{c|c|c}
    \hline
    \textbf{Category} & \textbf{Number of Images} & \textbf{Image Size} \\
    \hline
    Normal            & 100                       & 2048 × 1536         \\
     \hline
    Benign            & 100                       & 2048 × 1536         \\
     \hline
    In-situ carcinoma & 100                       & 2048 × 1536         \\
     \hline
    Invasive carcinoma& 100                       & 2048 × 1536         \\
    \hline
    Total  & 400             & 2048 × 1536 \\
    \hline
    \end{tabular}
\end{table}

 \begin{table}[H]
    \centering
    \caption{Detailed Information of the Mini-DDSM Dataset}
    \label{tab:mini_ddsm}
    \begin{tabular}{l|c|l}
    \hline
    \textbf{Category}       & \textbf{Number of Images} & \textbf{Image Size / Format} \\
    \hline
   
    Normal Images           & 500        & Variable                    \\
    \hline
    Benign Images           & 700        & Variable                    \\
    \hline
    Malignant Images        & 800        & Variable                    \\
    \hline
    Total Images            &  2000       & Variable                    \\
    \hline
    \end{tabular}
\end{table}

\subsection{Data Preprocessing and Splitting} 

All breast cancer histopathology images are initially resized to 224x224 pixels to match the input specifications for convolutional neural networks. Pixel values are normalized to the range [-1, 1] to expedite training and enhance generalization. Data augmentation techniques, such as random rotation and horizontal flipping, improve the model’s robustness by simulating real-world image variations. The dataset is then split into training, Validation, and test sets using a 6:1:3 ratio, as shown in Table 6. The training set is used for learning parameters, the Validation set is used for assessing and fine-tuning the model, and the test set is used for evaluating recognition and generalization performance. The training data is shuffled before model training to boost robustness and generalization further.

\begin{table*}[h]
    \centering
    \caption{ Number of Samples in Training, Validation, and Test Sets for Each Dataset}
    \label{Tal1}
    \begin{tabular}{c|c|c|c|c}
    \hline
    Dataset & Total Images & Training (60\%) & Validation (10\%) & Testing (30\%) \\
    \hline
    BreakHis 40X & 1995 & 1197 & 199 & 599 \\
    BreakHis 100X & 2081 & 1249 & 208 & 624  \\
    BreakHis 200X & 2013 & 1208 & 201 & 604  \\
    BreakHis 400X & 1820 & 1092 & 182 & 546  \\
    BACH & 400 & 240 & 40 & 120 \\
    Mini-DDSM & 2000 & 1200 & 200 & 600 \\
    \hline
    \end{tabular}
\end{table*}

\begin{table*}[t]
\centering
\caption{Performance of Network Models on BreakHis Dataset (Two-Class Classification)}
\label{your_table_label}
\scriptsize 
\begin{tabular}{cccccccccc}
\hline
\textbf{Network Models} & \multicolumn{4}{c}{\textbf{40X}} & \multicolumn{4}{c}{\textbf{100X}} \\ 
\cline{2-9}
 & \textbf{Accuracy} & \textbf{F1} & \textbf{IBA} & \textbf{GMean} & \textbf{Accuracy} & \textbf{F1} & \textbf{IBA} & \textbf{GMean} \\ 
\hline
SEResNet50 \cite{b39}& 0.883 & 0.779/0.883 & 0.860 & 0.860 & 0.850 & 0.720/0.849 & 0.813 & 0.812 \\ 
DenseNet121 \cite{b40} & 0.801 & 0.639/0.800 & 0.763 & 0.762 & 0.811 & 0.654/0.809 & 0.755 & 0.753 \\ 
VGG16 \cite{b41} & 0.902 & 0.813/0.902 & 0.877 & 0.877 & 0.850 & 0.724/0.851 & 0.830 & 0.829 \\ 
VGG16Inception \cite{b42}& 0.977 & 0.955/0.977 & 0.965 & 0.965 & 0.971 & 0.943/0.971 & 0.964 & 0.964 \\ 
ViT \cite{b43} & 0.835 & 0.693/0.833 & 0.769 & 0.767 & 0.834 & 0.694/0.833 & 0.788 & 0.787 \\ 
Swin-Transformer \cite{b44} & 0.843 & 0.711/0.844 & 0.819 & 0.818 & 0.877 & 0.767/0.876 & 0.836 & 0.835 \\ 
Dinov2\_Vitb14 \cite{b45} & 0.873 & 0.762/0.873 & 0.843 & 0.843 & 0.869 & 0.753/0.868 & 0.835 & 0.834 \\ 
ResNet50 \cite{b9}& 0.927 & 0.860/0.927 & 0.914 & 0.914 & 0.944 & 0.891/0.944 & 0.922 & 0.922 \\ 
\textbf{DALAResNet50} & \textbf{0.985} & \textbf{0.970/0.985} & \textbf{0.980} & \textbf{0.980} & \textbf{0.987} & \textbf{0.974/0.987} & \textbf{0.985} & \textbf{0.985} \\ 
\hline
\end{tabular}

\vspace{0.3cm} 

\begin{tabular}{cccccccccc}
\hline
\textbf{Network Models} & \multicolumn{4}{c}{\textbf{200X}} & \multicolumn{4}{c}{\textbf{400X}} \\ 
\cline{2-9}
 & \textbf{Accuracy} & \textbf{F1} & \textbf{IBA} & \textbf{GMean} & \textbf{Accuracy} & \textbf{F1} & \textbf{IBA} & \textbf{GMean} \\ 
\hline
SEResNet50 \cite{b39}& 0.906 & 0.820/0.906 & 0.876 & 0.875 & 0.876 & 0.770/0.878 & 0.864 & 0.863 \\ 
DenseNet121 \cite{b40}& 0.812 & 0.659/0.812 & 0.775 & 0.774 & 0.779 & 0.603/0.777 & 0.737 & 0.736 \\ 
VGG16 \cite{b41}& 0.851 & 0.722/0.850 & 0.812 & 0.811 & 0.848 & 0.718/0.848 & 0.817 & 0.817 \\ 
VGG16Inception \cite{b42}& 0.954 & 0.910/0.954 & 0.938 & 0.938 & 0.889 & 0.788/0.888 & 0.857 & 0.856 \\ 
ViT \cite{b43}& 0.841 & 0.705/0.840 & 0.801 & 0.800 & 0.832 & 0.691/0.831 & 0.803 & 0.803 \\ 
Swin-Transformer \cite{b44} & 0.876 & 0.766/0.875 & 0.838 & 0.837 & 0.846 & 0.724/0.851 & 0.791 & 0.789 \\ 
Dinov2\_Vitb14 \cite{b45} & 0.853 & 0.726/0.852 & 0.802 & 0.800 & 0.885 & 0.782/0.885 & 0.840 & 0.839 \\ 
ResNet50 \cite{b9}& 0.960 & 0.922/0.960 & 0.947 & 0.947 & 0.925 & 0.856/0.925 & 0.918 & 0.918 \\ 
\textbf{DALAResNet50} & \textbf{0.979} & \textbf{0.958/0.979} & \textbf{0.980} & \textbf{0.980} & \textbf{0.943} & \textbf{0.891/0.944} & \textbf{0.945} & \textbf{0.945} \\ 
\hline
\end{tabular}
\end{table*}

\begin{table*}[h]
\centering
\caption{Performance of Network Models on BACH Dataset (Four-Class Classification) and Mini-DDSM Dataset (Three-Class Classification)}
\label{table:combined_performance}
\scriptsize 
\renewcommand{\arraystretch}{1.2} 
\begin{tabular}{ccccccccc}
\hline
\textbf{Network Models} & \multicolumn{4}{c}{\textbf{BACH Dataset (Four-Class Classification)}} & \multicolumn{4}{c}{\textbf{Mini-DDSM Dataset (Three-Class Classification)}} \\ \cline{2-9}
 & \textbf{Accuracy} & \textbf{F1} & \textbf{IBA} & \textbf{GMean} & \textbf{Accuracy} & \textbf{F1} & \textbf{IBA} & \textbf{GMean} \\ \hline
SEResNet50 \cite{b39} & 0.770 & 0.743/0.77 & 0.583 & 0.553 & 0.7006 & 0.699/0.701 & 0.709 & 0.768 \\ 
DensNet121 \cite{b40}& 0.703 & 0.680/0.703 & 0.458 & 0.438 & 0.6301 & 0.627/0.63 & 0.642 & 0.716 \\ 
VGG16 \cite{b41}& 0.788 & 0.781/0.788 & 0.652 & 0.637 & 0.6607 & 0.66/0.661 & 0.672 & 0.736 \\ 
VGG16Inception \cite{b42} & 0.727 & 0.864/0.727 & 0.650 & 0.654 & 0.7247 & 0.728/0.725 & 0.728 & 0.786 \\ 
ViT \cite{b43} & 0.746 & 0.734/0.745 & 0.580 & 0.555 & 0.7515 & 0.72/0.752 & 0.580 & 0.555 \\ 
Swin-Transformer \cite{b44} & 0.748 & 0.556/0.745 & 0.500 & 0.436 & 0.5110 & 0.51/0.511 & 0.52 & 0.619 \\ 
Dinov2\_Vitb14 \cite{b45} & 0.856 & 0.730/0.855 & 0.500 & 0.353 & 0.667 & 0.665/0.667 & 0.678 & 0.742 \\ 
ResNet50 \cite{b9} & 0.855 & 0.838/0.855 & 0.664 & 0.635 & 0.6686 & 0.672/0.669 & 0.677 & 0.74 \\ 
\textbf{DALAResNet50} & \textbf{0.891} & \textbf{0.891/0.885} & \textbf{0.748} & \textbf{0.734} & \textbf{0.8070} & \textbf{0.806/0.807} & \textbf{0.818} & \textbf{0.851} \\ \hline
\end{tabular}
\end{table*}

\subsection{ Metrics}

To assess our model’s performance holistically, we used four metrics: Accuracy \cite{b46,b47,b48}, F1 Score \cite{b49,b52} Imbalance Balanced Accuracy (IBA) \cite{b50,b53}, and Geometric Mean (GMean) \cite{b51,b54}. Accuracy measures the proportion of correctly predicted samples. The F1 Score combines precision and recall, crucial for imbalanced datasets. IBA adjusts accuracy to fairly evaluate both majority and minority classes. GMean calculates the geometric mean of class accuracies, emphasizing performance on less represented classes.

\subsection{Comparison with Other Methods}

In this section, we compare our proposed model with several state-of-the-art methods: SEResNet50  \cite{b39}, DenseNet-121 \cite{b40}, VGG16 \cite{b41}, VGG16Inception \cite{b42}, ViT \cite{b43}, Swin-Transformer \cite{b44}, and Dinov2\_Vitb14 \cite{b45}. Tables 7 and 8 show our model’s superior performance on the BreakHis, BACH, and Mini-DDSM datasets regarding accuracy, F1 score, IBA, and GMean. Additionally, as part of the ablation study, we evaluated the model’s performance across different resolutions (40x, 100x, 200x, and 400x) on the BreakHis dataset. As shown in the confusion matrices in Fig.6, DALAResNet50 consistently outperformed the standard ResNet50 model across all magnification levels, demonstrating that the newly added lightweight dual-activation attention module significantly enhances the model’s performance.

Tables 9 and 10 provide a detailed analysis of the parameter counts and convergence speeds. Although DenseNet-121 has the fewest parameters, DALAResNet50 achieved the shortest convergence time in the evaluation, further validating its efficiency in medical image classification tasks. Additionally, Tables  11 and 12 present DALAResNet50’s performance across all magnification levels in terms of sensitivity, specificity, positive predictive value (PPV), and negative predictive value (NPV). These results demonstrate that DALAResNet50 not only effectively reduces false positives and false negatives but also significantly improves the accuracy in detecting both positive and negative samples through its higher sensitivity and specificity. These features are particularly important for addressing the challenges posed by imbalanced breast cancer datasets.

\begin{table*}[t]
    \centering
    \caption{Parameters Count and Convergence Time of Network Models on BreakHis Dataset}
    \label{Tal5}
    \begin{tabular}{m{110pt}<{\centering} cccc|c|c}
         \hline
         \multirow{2}{*}{\textbf{Network Models}} & \multicolumn{4}{c|}{\textbf{Convergence Time (seconds)}} & \multirow{2}{*}{\textbf{Total Parameters}} & \multirow{2}{*}{\textbf{Average Time}} \\
         \cline{2-5}
         & \textbf{40X} & \textbf{100X} & \textbf{200X} & \textbf{400X} & & \\
         \hline
         SEResNet50  \cite{b39} & 2696.22 & 2814.30 & 2711.89 & 2356.18 & 27073090 & 2644.65 \\
         DenseNet121  \cite{b40}& 2352.46 & 2643.79 & 2423.36 & 2078.34 & \textbf{9028970} & 2347.49 \\
         VGG16  \cite{b41}& 3212.95 & 3182.54 & 3011.53 & 2640.19 & 134268738 & 3011.80 \\
         VGG16Inception  \cite{b42}& 3074.99 & 3253.50 & 3303.12 & 2682.61 & 38139938 & 3078.56 \\
         ViT  \cite{b43}& 2932.11 & 3022.54 & 3013.86 & 2691.16 & 51375106 & 2914.92 \\
         Swin-Transformer  \cite{b44}& 3020.11 & 3132.42 & 2940.98 & 2592.96 & 86745274 & 2921.62 \\
         Dinov2\_Vitb14  \cite{b45}& 2895.02 & 2303.25 & 2245.98 & 2017.15 & 86582018 & 2365.35 \\
         ResNet50  \cite{b9}& 2920.53 & 2933.26 & 2307.94 & 2097.33 & 21797672 & 2564.77 \\
         \textbf{DALAResNet50 } & \textbf{2133.02} & \textbf{2319.50} & \textbf{1990.58} & \textbf{1896.82} & 25611842 & \textbf{2084.98} \\
         \hline
         
    \end{tabular}
\end{table*}

\begin{table*}[t]
    \centering
    \caption{Parameters Count and Convergence Time of Network Models on BACH and Mini-DDSM Datasets}
    \label{tab:combined_performance_metrics}
    \begin{tabular}{l|c|c|c|c}
    \hline
    \textbf{Network Models} & \multicolumn{2}{c|}{\textbf{BACH Dataset}} & \multicolumn{2}{c}{\textbf{Mini-DDSM Dataset}} \\
    \cline{2-5}
    & \textbf{Total Parameters} & \textbf{Average Time} & \textbf{Total Parameters} & \textbf{Average Time} \\
    \hline
    SEResNet50 \cite{b39} & 27074116 & 2921.67 & 27073603 & 7272.75 \\
    DensNet121 \cite{b40} & \textbf{9029996} & 3275.27 & \textbf{9029483} & 6384.11 \\
    VGG16 \cite{b41} & 134276934 & 3303.98 & 134272835 & 7589.96 \\
    VGG16Inception \cite{b42} & 38148134 & 3432.98 & 38144035 & 8089.23 \\
    ViT \cite{b43} & 51385352 & 2872.36 & 51376131 & 6555.35 \\
    Swin-Transformer \cite{b44} & 86753470 & 3488.72 & 86746299 & 9715.44 \\
    Dinov2\_Vitb14 \cite{b45} & 86583556 & 2966.12 & 86582787 & 6432.41 \\
    ResNet50 \cite{b9} & 21805868 & 3723.49 & 21797672 & 8268.68 \\
    \textbf{DALAResNet50} & 24042692 & \textbf{2475.49} & 24040643 & \textbf{5990.31} \\
    \hline
    \end{tabular}
\end{table*}

\begin{figure*}[t]
    \centering
    
    \begin{minipage}{0.22\textwidth}
        \centering
        \includegraphics[width=\linewidth]{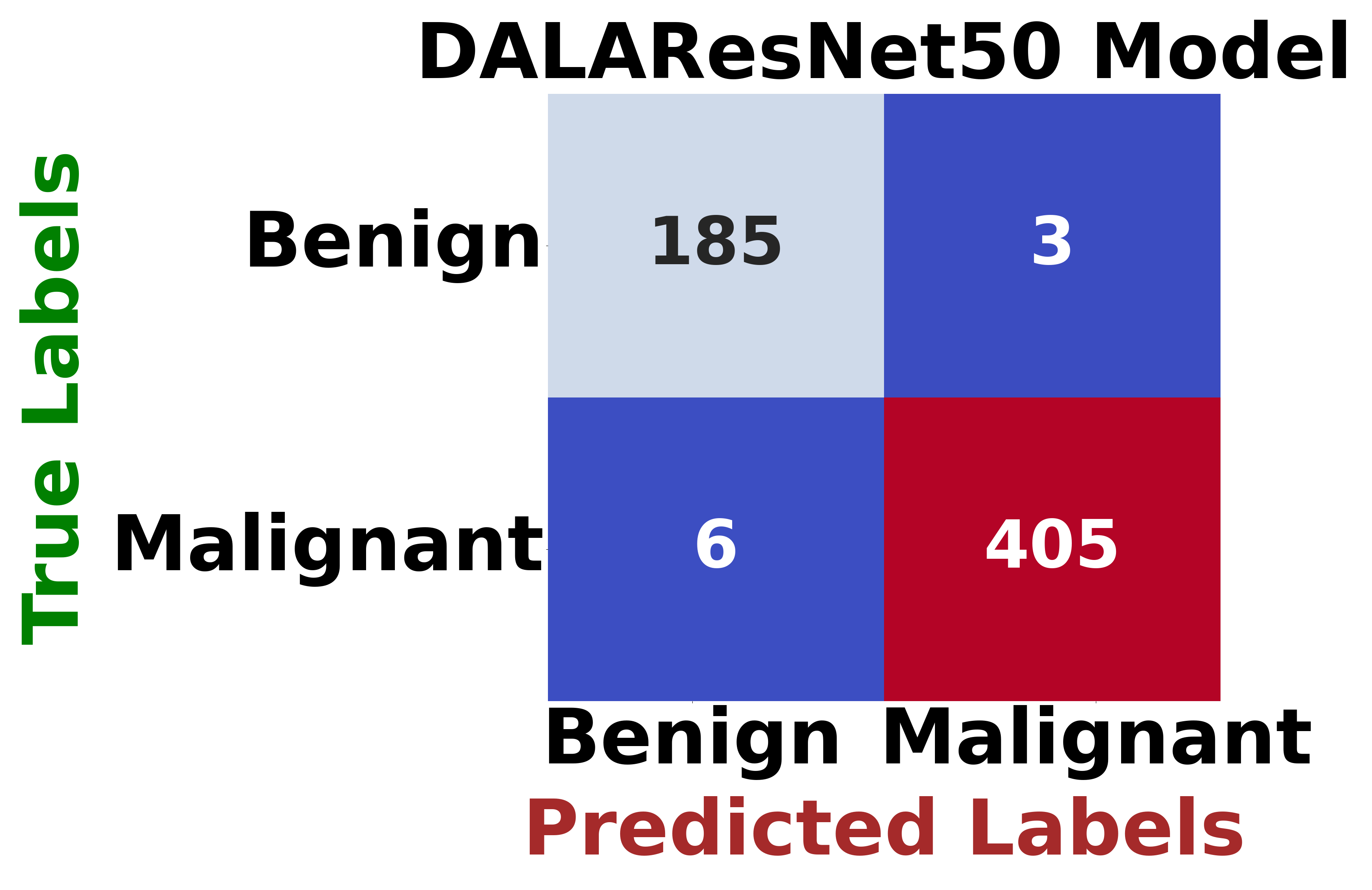}
    \end{minipage}%
    \hspace{0.02\textwidth}
    \begin{minipage}{0.22\textwidth}
        \centering
        \includegraphics[width=\linewidth]{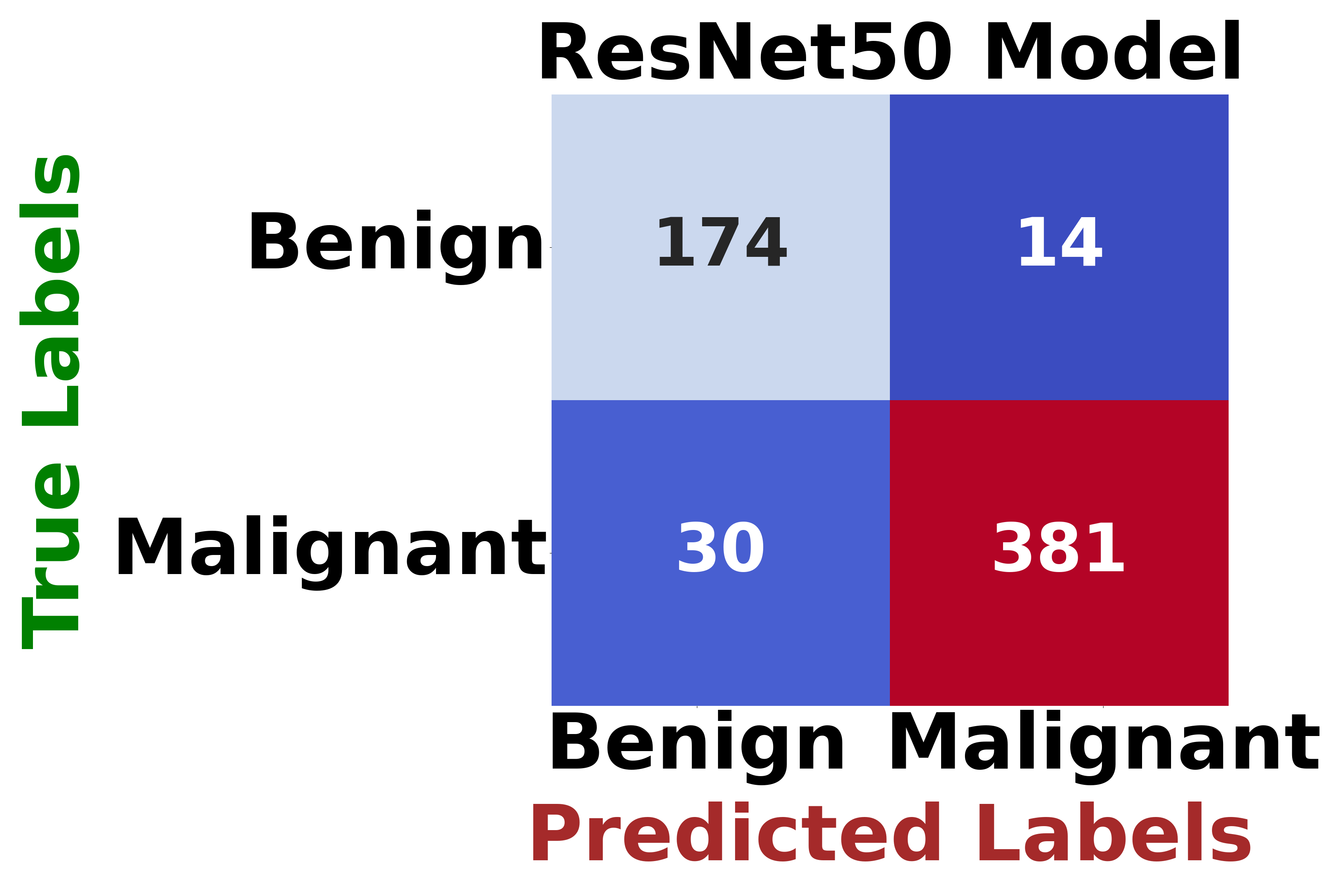}
    \end{minipage}%
    \hspace{0.05\textwidth}
    \begin{minipage}{0.22\textwidth}
        \centering
        \includegraphics[width=\linewidth]{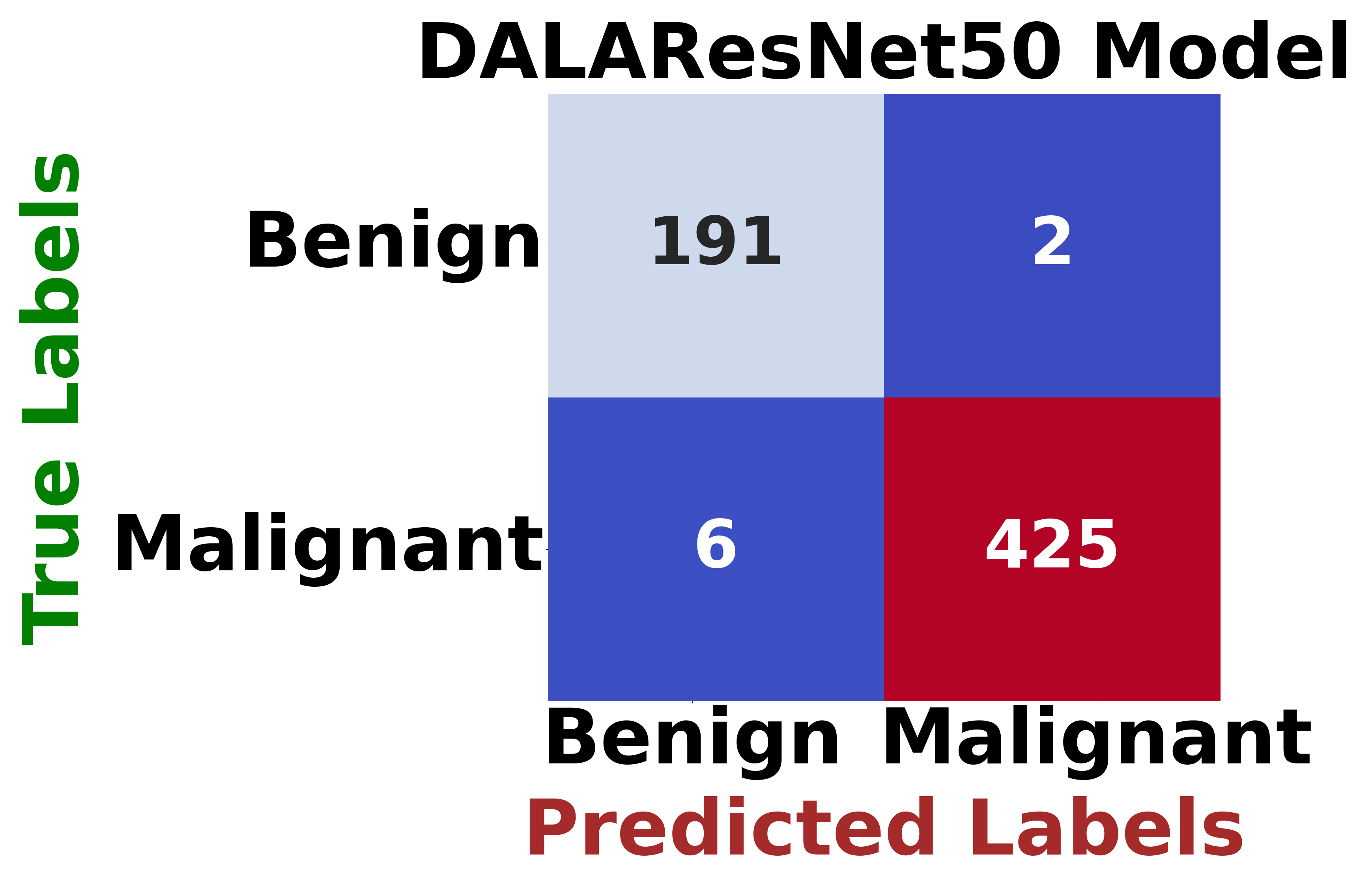}
    \end{minipage}%
    \hspace{0.02\textwidth}
    \begin{minipage}{0.22\textwidth}
        \centering
        \includegraphics[width=\linewidth]{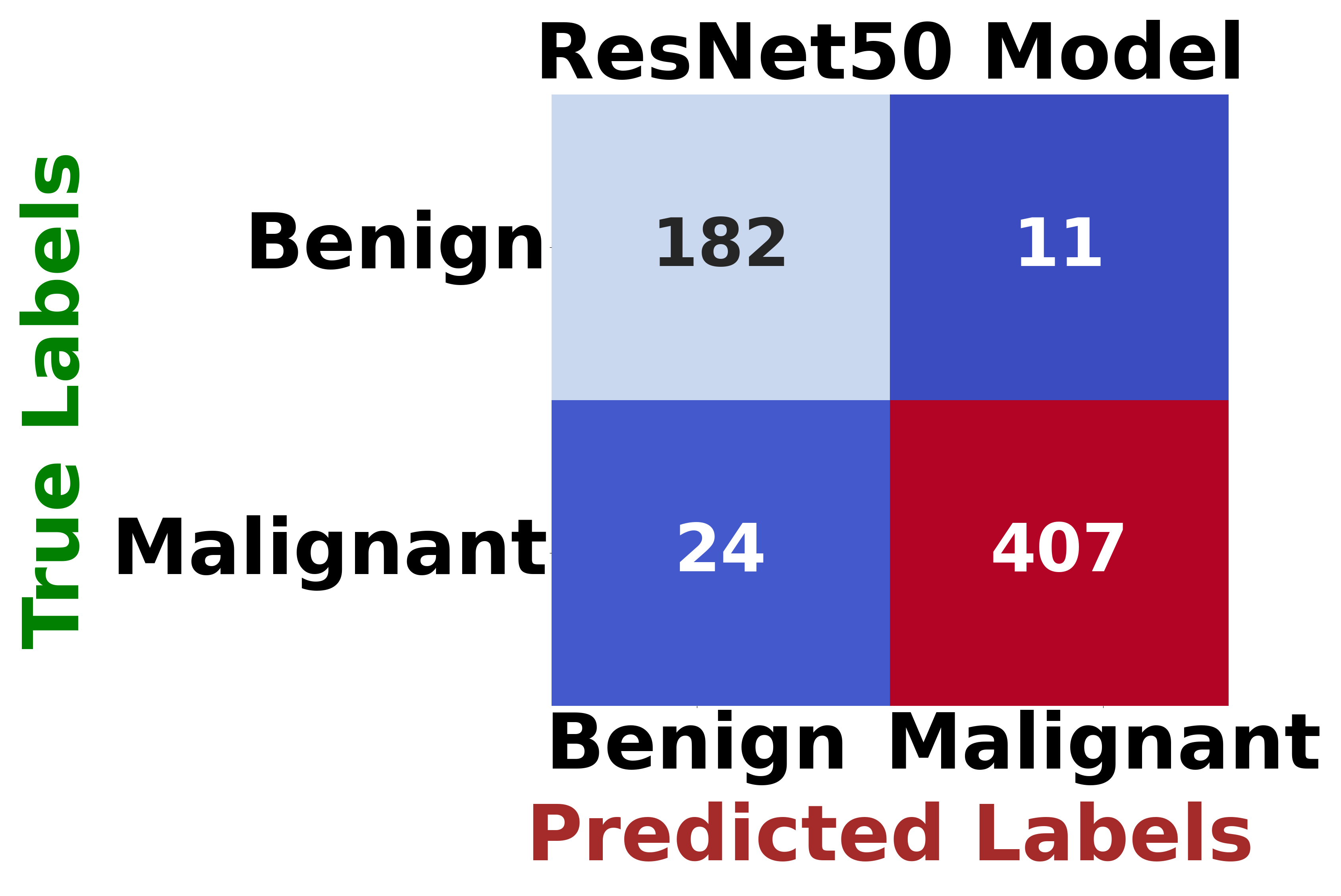}
    \end{minipage}

    \vspace{0.3cm} 
    \begin{minipage}{0.22\textwidth}
        \centering
        40x
    \end{minipage}%
    \hspace{0.29\textwidth} 
    \begin{minipage}{0.22\textwidth}
        \centering
        100x
    \end{minipage}

    \vspace{0.5cm} 

    \begin{minipage}{0.22\textwidth}
        \centering
        \includegraphics[width=\linewidth]{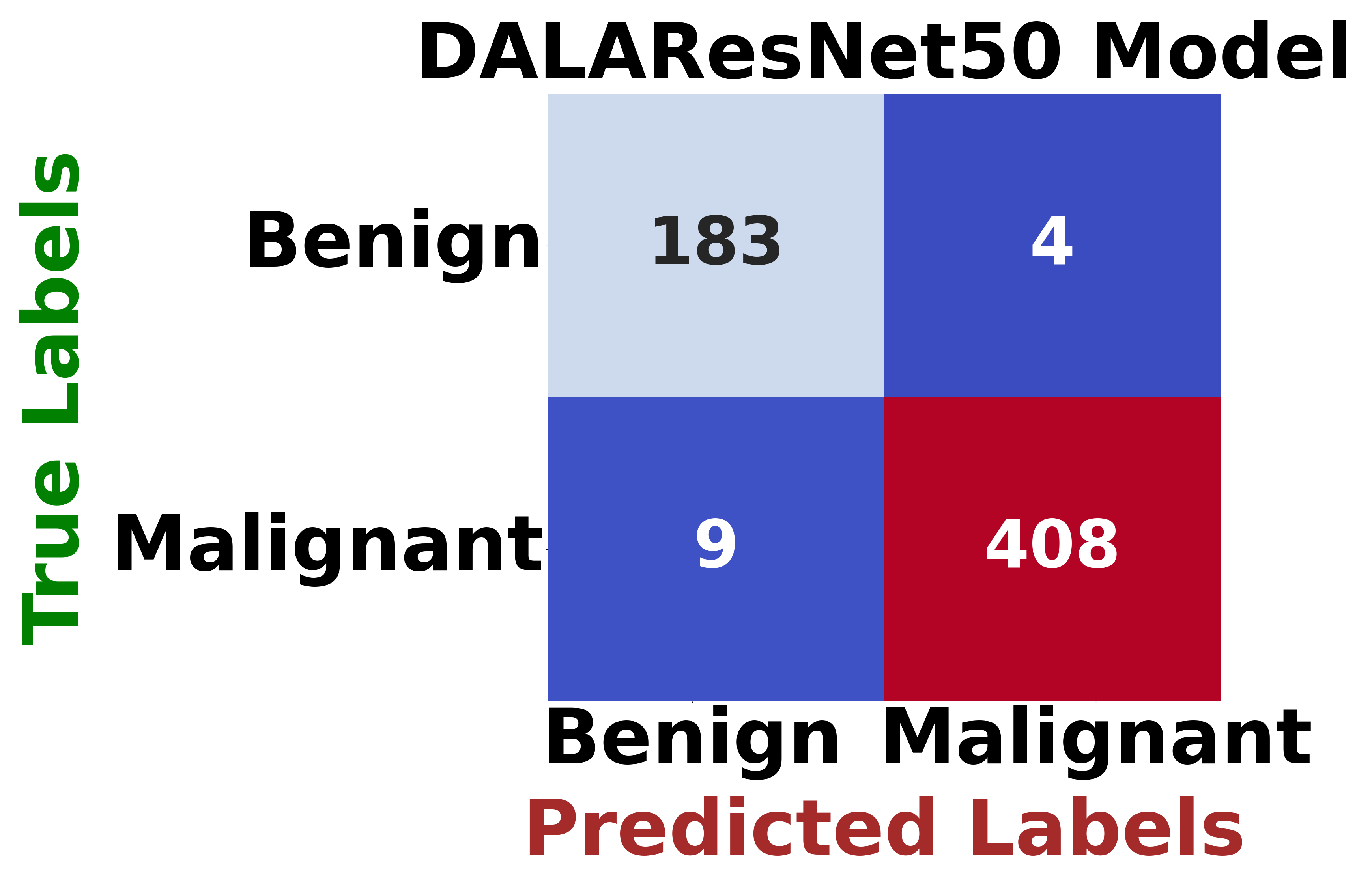}
    \end{minipage}%
    \hspace{0.02\textwidth}
    \begin{minipage}{0.22\textwidth}
        \centering
        \includegraphics[width=\linewidth]{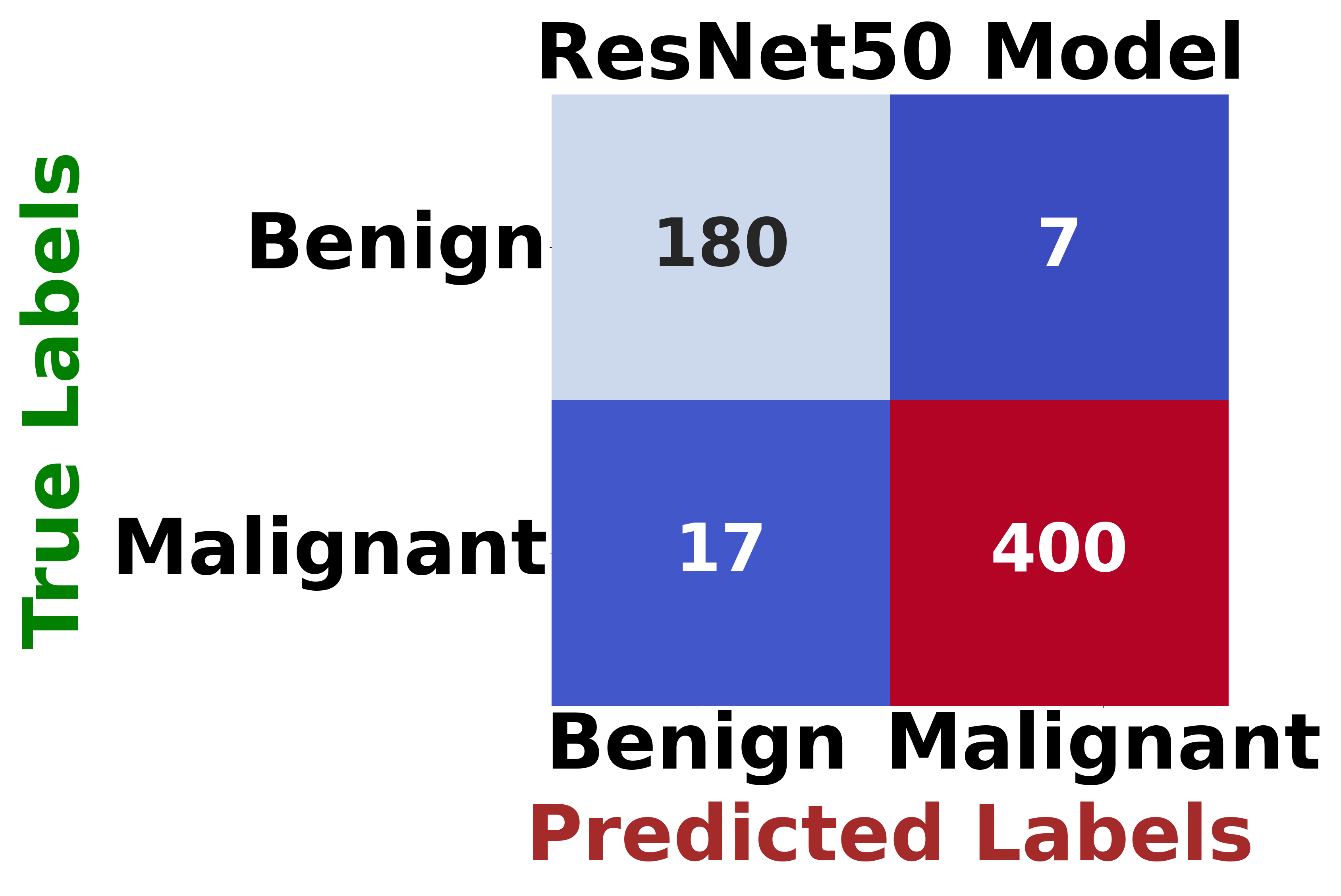}
    \end{minipage}%
    \hspace{0.05\textwidth}
    \begin{minipage}{0.22\textwidth}
        \centering
        \includegraphics[width=\linewidth]{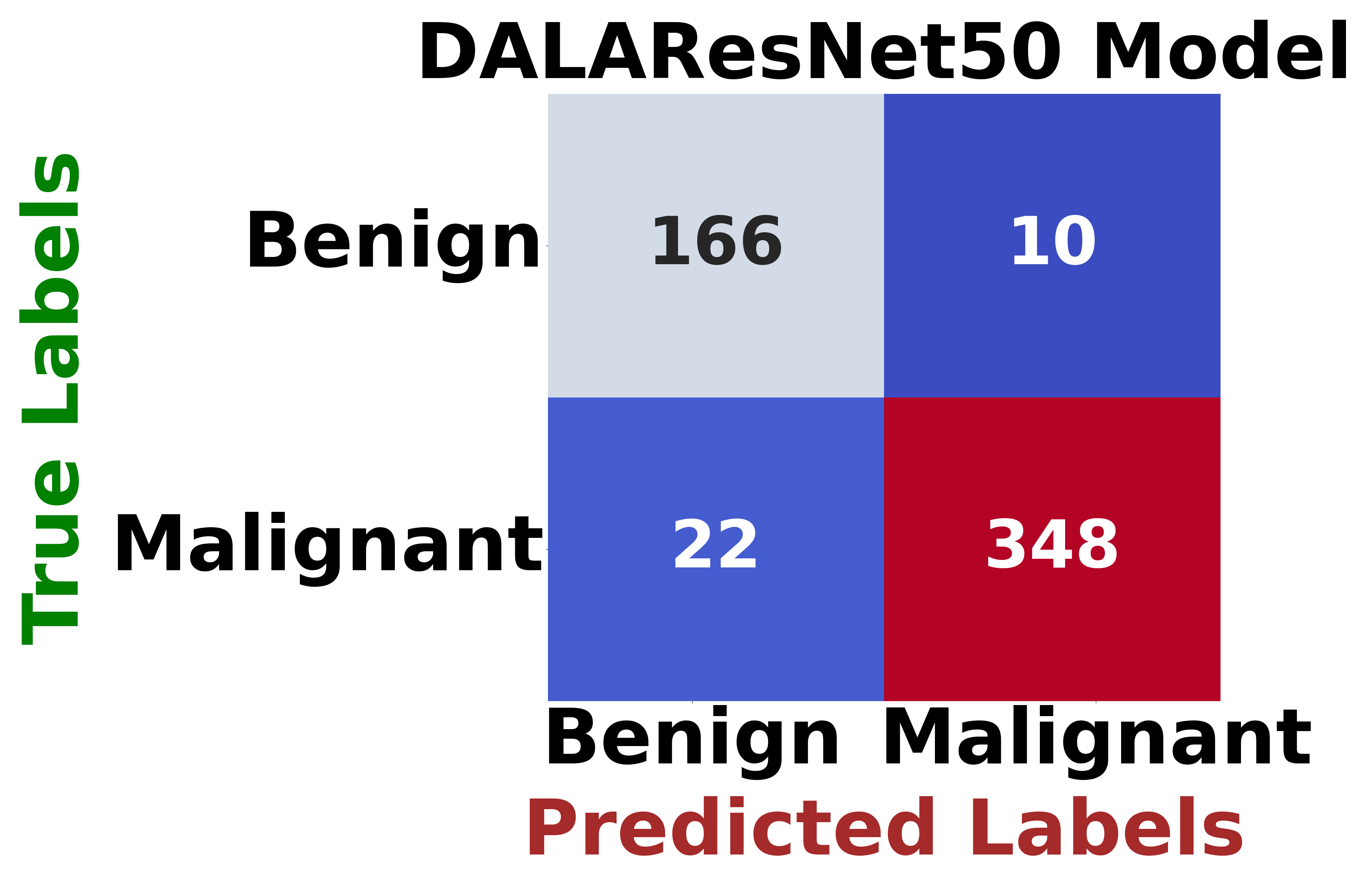}
    \end{minipage}%
    \hspace{0.02\textwidth}
    \begin{minipage}{0.22\textwidth}
        \centering
        \includegraphics[width=\linewidth]{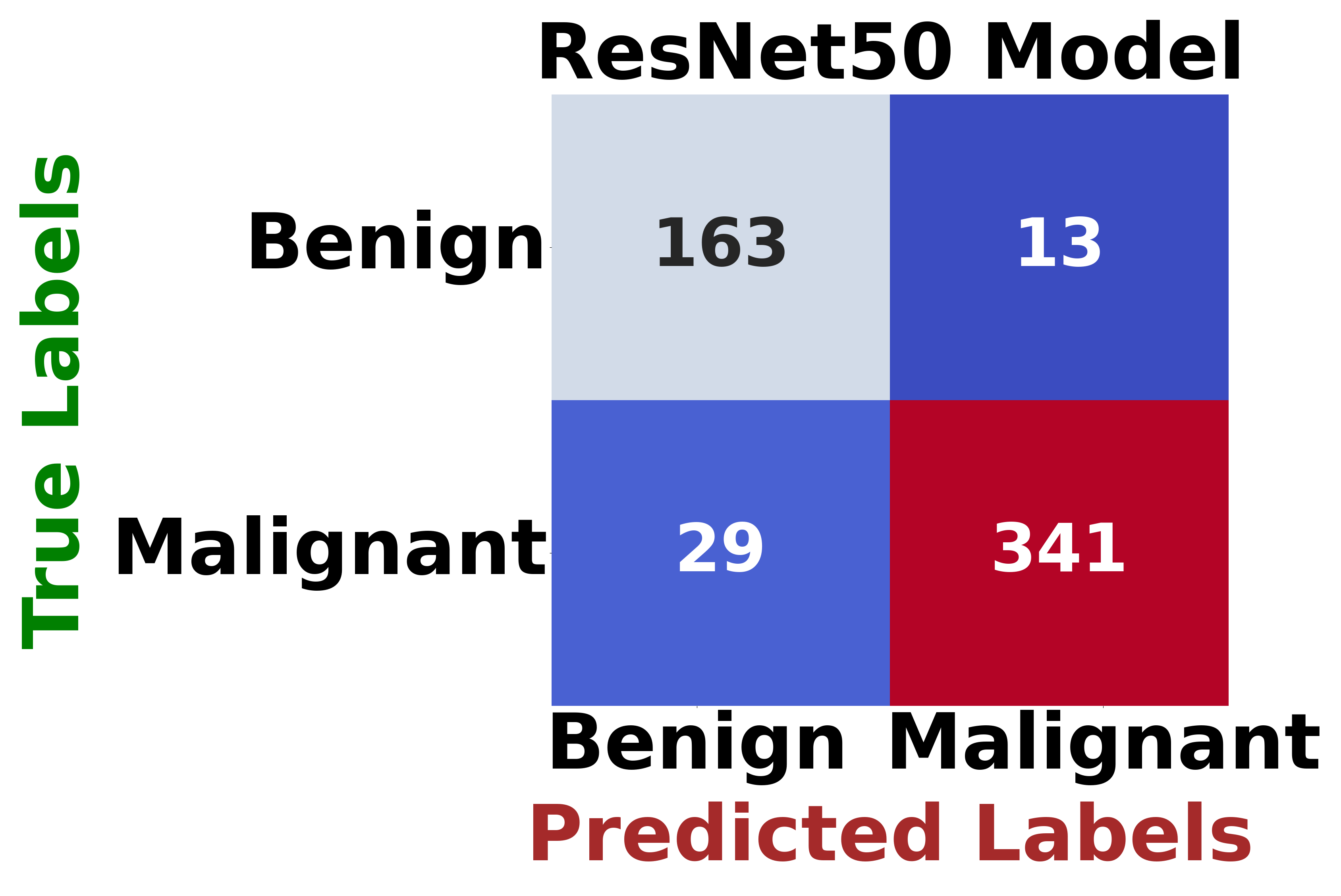}
    \end{minipage}

    \vspace{0.3cm} 
    \begin{minipage}{0.22\textwidth}
        \centering
        200x
    \end{minipage}%
    \hspace{0.29\textwidth} 
    \begin{minipage}{0.22\textwidth}
        \centering
        400x
    \end{minipage}

    \caption{Comparison of the confusion matrices of the DALAResNet50 model and the standard ResNet50 model.}
\end{figure*}

\begin{table*}[t]
\centering
\caption{Sensitivity, Specificity, PPV, and NPV of Network Models on the Imbalanced BreakHis Dataset }
\label{your_table_label}
\scriptsize 
\begin{tabular}{cccccccccc}
\hline
\textbf{Network Models} & \multicolumn{4}{c}{\textbf{40X}} & \multicolumn{4}{c}{\textbf{100X}} \\ 
\cline{2-9}
 & \textbf{Sensitivity} & \textbf{Specificity} & \textbf{PPV} & \textbf{NPV} & \textbf{Sensitivity} & \textbf{Specificity} & \textbf{PPV} & \textbf{NPV} \\ 
\hline
SEResNet50 \cite{b39}& 0.883 & 0.883 & 0.943 & 0.776 &  0.849 & 0.850 & 0.927 &  0.716 \\ 
DenseNet121 \cite{b40} &  0.800 & 0.803 & 0.899 & 0.648 & 0.810 & 0.813 & 0.906 & 0.657 \\ 
VGG16 \cite{b41} &  0.903 & 0.899 & 0.951 & 0.809 & 0.849 & 0.850 & 0.927 &  0.716\\ 
VGG16Inception \cite{b42}& 0.976 & 0.979 & 0.990 & 0.948 & 0.972 & 0.969 & 0.986 & 0.940 \\ 
ViT \cite{b43} & 0.835 & 0.835 & 0.917& 0.698 & 0.833 & 0.834 & 0.918 & 0.691 \\ 
Swin-Transformer \cite{b44} &  0.844 & 0.840 & 0.920 &0.712 & 0.877 & 0.876 &  0.940 & 0.761 \\ 
Dinov2\_Vitb14 \cite{b45} & 0.873 & 0.872 &  0.937 & 0.759 & 0.868 & 0.870 & 0.937 & 0.747 \\ 
ResNet50 \cite{b9}& 0.927 & 0.926 & 0.965 & 0.853 & 0.944 & 0.943 & 0.974 & 0.883 \\ 
\textbf{DALAResNet50} & \textbf{0.985} & \textbf{0.984} & \textbf{0.993} & \textbf{0.969} & \textbf{0.986} & \textbf{0.990} & \textbf{0.995} & \textbf{0.970} \\ 
\hline
\end{tabular}

\vspace{0.3cm} 

\begin{tabular}{cccccccccc}
\hline
\textbf{Network Models} & \multicolumn{4}{c}{\textbf{200X}} & \multicolumn{4}{c}{\textbf{400X}} \\ 
\cline{2-9}
 & \textbf{Sensitivity} & \textbf{Specificity} & \textbf{PPV} & \textbf{NPV} & \textbf{Sensitivity} & \textbf{Specificity} & \textbf{PPV} & \textbf{NPV} \\ 
\hline
SEResNet50 \cite{b39}& 0.906 & 0.904 & 0.955 & 0.813 & 0.876 & 0.875 & 0.936 & 0.770 \\ 
DenseNet121 \cite{b40}& 0.811 & 0.813 & 0.906 & 0.658 & 0.780 &  0.778& 0.881 & 0.628 \\ 
VGG16 \cite{b41}& 0.851 & 0.850 &  0.927& 0.719 & 0.848 & 0.847 & 0.921 & 0.727 \\ 
VGG16Inception \cite{b42}& 0.954 & 0.952 & 0.978 &  0.904 & 0.889 & 0.892 & 0.945 & 0.793 \\ 
ViT \cite{b43}& 0.842 & 0.840 & 0.921 &  0.704 & 0.832 & 0.830 & 0.911 & 0.702 \\ 
Swin-Transformer \cite{b44} & 0.875 & 0.877 & 0.941 & 0.759 & 0.846 & 0.847 & 0.920 & 0.723 \\ 
Dinov2\_Vitb14 \cite{b45} & 0.854 & 0.850 &  0.927 & 0.723 &  0.883 & 0.886 & 0.942 & 0.784\\ 
ResNet50 \cite{b9}& 0.959 & 0.963 & 0.983 & 0.914 & 0.924 & 0.926 & 0.963 & 0.853\\ 
\textbf{DALAResNet50} & \textbf{0.978} & \textbf{0.979} & \textbf{0.990} & \textbf{0.953} & \textbf{0.943} & \textbf{0.943} & \textbf{0.972} & \textbf{0.888} \\ 
\hline
\end{tabular}
\end{table*}

\begin{table*}[h]
\centering
\caption{Sensitivity, Specificity, PPV, and NPV of Network Models on the Imbalanced BACH Dataset  and Mini-DDSM Dataset }
\label{table:combined_performance}
\scriptsize 
\renewcommand{\arraystretch}{1.2} 
\begin{tabular}{ccccccccc}
\hline
\textbf{Network Models} & \multicolumn{4}{c}{\textbf{BACH Dataset (Four-Class Classification)}} & \multicolumn{4}{c}{\textbf{Mini-DDSM Dataset (Three-Class Classification)}} \\ \cline{2-9}
 & \textbf{Sensitivity } & \textbf{Specificity} & \textbf{PPV} & \textbf{NPV} & \textbf{Sensitivity} & \textbf{Specificity} & \textbf{PPV} & \textbf{NPV} \\ \hline
SEResNet50 \cite{b39} &  0.771 & 0.768 & 0.909 & 0.527 & 0.700 & 0.700 & 0.806 & 0.568 \\ 
DensNet121 \cite{b40}& 0.703 & 0.704 & 0.877 & 0.441 & 0.631 & 0.628 & 0.751 & 0.489 \\ 
VGG16 \cite{b41}& 0.781 & 0.780 & 0.914 & 0.543 &  0.659 & 0.661 & 0.776 & 0.522 \\ 
VGG16Inception \cite{b42} & 0.727 & 0.728 &  0.889 & 0.470 & 0.725 & 0.722 &  0.823& 0.596 \\ 
ViT \cite{b43} & 0.747& 0.744& 0.897&0.495 & 0.753 & 0.750 & 0.843 &  0.631 \\ 
Swin-Transformer \cite{b44} & 0.748 & 0.748 & 0.899 & 0.497 & 0.513 & 0.511&  0.651 & 0.371 \\ 
Dinov2\_Vitb14 \cite{b45} & 0.856&  0.856 & 0.947 & 0.665 & 0.669 & 0.667 & 0.781 & 0.531 \\ 
ResNet50 \cite{b9} & 0.855 & 0.856 & 0.947& 0.663 & 0.669 & 0.667 & 0.781 & 0.531 \\ 
\textbf{DALAResNet50} & \textbf{0.891} & \textbf{0.892} & \textbf{0.961} & \textbf{0.731} & \textbf{0.809} & \textbf{0.806} & \textbf{0.881} & \textbf{0.704} \\ \hline
\end{tabular}
\end{table*}

\section{DT Grad-CAM}

The interpretability of deep learning models is crucial, particularly in areas such as breast cancer diagnosis. Grad-CAM is a popular technique for visualizing class-specific regions of interest in images. This article introduces an enhancement called DT Grad-CAM (Dynamic Threshold Grad-CAM), which improves Grad-CAM by incorporating noise robustness, weighted averaging, adaptive thresholding using the Otsu method, and morphological operations. These enhancements collectively improve the clarity and relevance of the visualizations.

\begin{table*}[t]
\centering
\scriptsize
\caption{Comparison of DT Grad-CAM and Grad-CAM}
\begin{tabular}{p{4cm}p{4cm}p{4cm}}
\hline
\textbf{Comparison Aspect} & \textbf{DT Grad-CAM} & \textbf{Grad-CAM} \\
\hline
Adaptive Thresholding & Utilizes optimal thresholding. & Lacks adaptive thresholding. \\

Visualization Clarity & Clear and focused. & Dispersed and less interpretable. \\

Interpretability & Sharp and meaningful highlights. & Basic interpretability. \\

Noise Reduction & Effectively minimizes noise. & Retains higher noise levels. \\

Diagnostic Relevance & Highly precise for diagnostics. & Less precise, reducing accuracy. \\
\hline
\end{tabular}
\end{table*}

\subsection{DT Grad-CAM Method}

The DT Grad-CAM method enhances the traditional Grad-CAM approach through several steps. Each step improves the visualization of class-specific regions, making the results more interpretable and relevant for diagnostic purposes.

\subsection*{1. Noisy Input Generation}

To improve robustness and capture more informative regions, Gaussian noise is added to the input image multiple times. For each iteration \( i \) out of \( N \) total iterations, the noisy input is generated as:

\begin{equation}
x_{\text{noisy}}^{(i)} = x + \sigma \cdot \mathcal{N}(0, 1)
\end{equation}

where:

\begin{itemize}
    \item \( x \) is the original input image.
    \item \( \sigma \) is the noise level parameter.
    \item \( \mathcal{N}(0, 1) \) represents a Gaussian distribution with mean 0 and standard deviation 1.
\end{itemize}

This step introduces variability, allowing the model to explore different perturbations and enhancing the robustness of the Grad-CAM visualization.

\subsection*{2. Grad-CAM Calculation for Noisy Inputs}

For each noisy input \( x_{\text{noisy}}^{(i)} \), the Grad-CAM heatmap is computed:

\begin{equation}
L_{c}^{\text{Grad-CAM}, (i)} = \text{ReLU} \left( \sum_k \alpha_k^{c, (i)} A^{k, (i)} \right)
\end{equation}

where:

\begin{itemize}
    \item \( A^{k, (i)} \) is the \( k \)-th feature map from the convolutional layer for the \( i \)-th noisy input.
    \item \( \alpha_k^{c, (i)} \) is the weight for feature map \( A^{k, (i)} \), computed as:

    \begin{equation}
    \alpha_k^{c, (i)} = \frac{1}{Z} \sum_{p} \sum_{q} \frac{\partial y^{c}}{\partial A_{pq}^{k, (i)}}
    \end{equation}

    \item \( y^{c} \) is the score for class \( c \).
    \item \( Z \) is the total number of elements in the feature map (i.e., width times height).
\end{itemize}

This step leverages gradient information from multiple noisy inputs to highlight important features contributing to the class prediction.

\subsection*{3. Weighted Averaging of Heatmaps}

The Grad-CAM heatmaps from all iterations are combined using a weighted average to emphasize more reliable maps:

\begin{equation}
L_{c}^{\text{Avg}} = \frac{\sum_{i=1}^{N} w_i L_{c}^{\text{Grad-CAM}, (i)}}{\sum_{i=1}^{N} w_i}
\end{equation}

where:

\begin{itemize}
    \item \( w_i \) are the weights assigned to each iteration, typically forming a decreasing sequence (e.g., linearly decreasing from 1 to 0.5).
    \item \( N \) is the total number of iterations.
\end{itemize}

This weighted averaging reduces the impact of noise and highlights consistent patterns across different perturbations.

\subsection*{4. Otsu Threshold Calculation}

An adaptive threshold \( T \) is computed using Otsu's method on the averaged heatmap \( L_{c}^{\text{Avg}} \):

\begin{equation}
\sigma_B^2(T) = \omega_1(T) \omega_2(T) [\mu_1(T) - \mu_2(T)]^2
\end{equation}

where:

\begin{itemize}
    \item \( \sigma_B^2(T) \) is the between-class variance for threshold \( T \).
    \item \( \omega_1(T) \) and \( \omega_2(T) \) are the probabilities of foreground and background pixels at threshold \( T \), respectively.
    \item \( \mu_1(T) \) and \( \mu_2(T) \) are the means of the foreground and background pixels at threshold \( T \), respectively.
\end{itemize}

By iterating through all possible thresholds \( T \), the one that maximizes \( \sigma_B^2(T) \) is selected as the Otsu threshold.

\begin{table*}[t]
\centering
\caption{Performance of CAM Methods Across Different Magnifications}
\label{your_table_label}
\scriptsize 
\begin{tabular}{cccccccccc}
\hline
\textbf{Method} & \multicolumn{4}{c}{\textbf{40X}} & \multicolumn{4}{c}{\textbf{100X}} \\
\cline{2-9}
& \textbf{IoU} & \textbf{Dice} & \textbf{Recall} & \textbf{F1 Score} & \textbf{IoU} & \textbf{Dice} & \textbf{Recall} & \textbf{F1 Score} \\
\hline
GradCAM \cite{b60}& 0.3064 & 0.4628 & 0.4746 & 0.4628 & 0.3132 & 0.4704 & 0.4758 & 0.4704 \\

ScoreCAM \cite{b61} & 0.3539 & 0.5122 & 0.5916 & 0.5122 & 0.3735 & 0.5334 & 0.6040 & 0.5334 \\ Eigen-CAM \cite{Muhammad2021} & 0.3735 & 0.5342 & 0.6332 & 0.5342 & 0.3864 & 0.5465 &0.6334 & 0.5465 \\
\textbf{DT-GradCAM} & \textbf{0.3993} & \textbf{0.5634} & \textbf{0.7143} &\textbf{ 0.5634 }& \textbf{0.4009} & \textbf{0.5644}& \textbf{0.6927} & \textbf{0.5644} \\
\hline
\end{tabular}

\vspace{0.3cm} 

\begin{tabular}{cccccccccc}
\hline
\textbf{Method} & \multicolumn{4}{c}{\textbf{200X}} & \multicolumn{4}{c}{\textbf{400X}} \\
\cline{2-9}
& \textbf{IoU} & \textbf{Dice} & \textbf{Recall} & \textbf{F1 Score} & \textbf{IoU} & \textbf{Dice} & \textbf{Recall} & \textbf{F1 Score} \\
\hline
GradCAM \cite{b60} & 0.2991 & 0.4542 & 0.4516 & 0.4542 & 0.2969 & 0.4508 & 0.4472 & 0.4508 \\

ScoreCAM \cite{b61} & 0.3713 & 0.5329 & 0.6184 & 0.5329 & 0.3651 & 0.5252 & 0.6050 & 0.5252 \\

EigenCAM \cite{Muhammad2021} & 0.3789 & 0.5388 & 0.6325 & 0.5388 & 0.3781 & 0.5379 & 0.6267 & 0.5379 \\

\textbf{DT-GradCAM }& \textbf{0.3931} & \textbf{0.5567} & \textbf{0.6901} & \textbf{0.5567} &\textbf{0.3940} & \textbf{0.5570} & \textbf{0.6888}& \textbf{0.5570}
\\
\hline
\end{tabular}
\end{table*}

\begin{figure*}[t]
\centering
\begin{minipage}{0.18\textwidth}
\centering
\includegraphics[width=\textwidth]{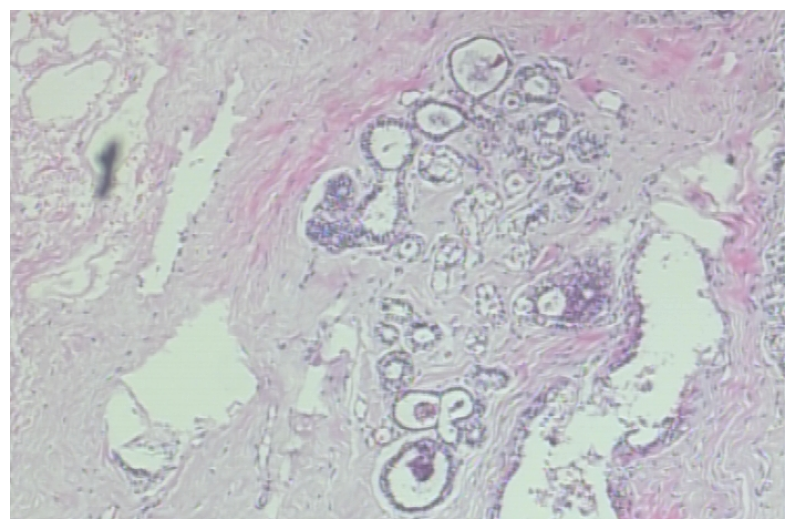}
\\40X-Original
\end{minipage}
\hfill
\begin{minipage}{0.18\textwidth}
\centering
\includegraphics[width=\textwidth]{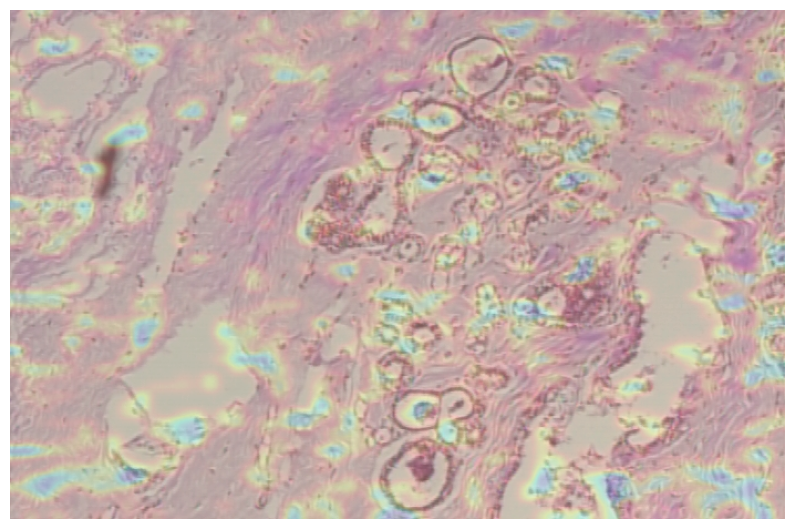}
\\40X-GradCAM
\end{minipage}
\hfill
\begin{minipage}{0.18\textwidth}
\centering
\includegraphics[width=\textwidth]{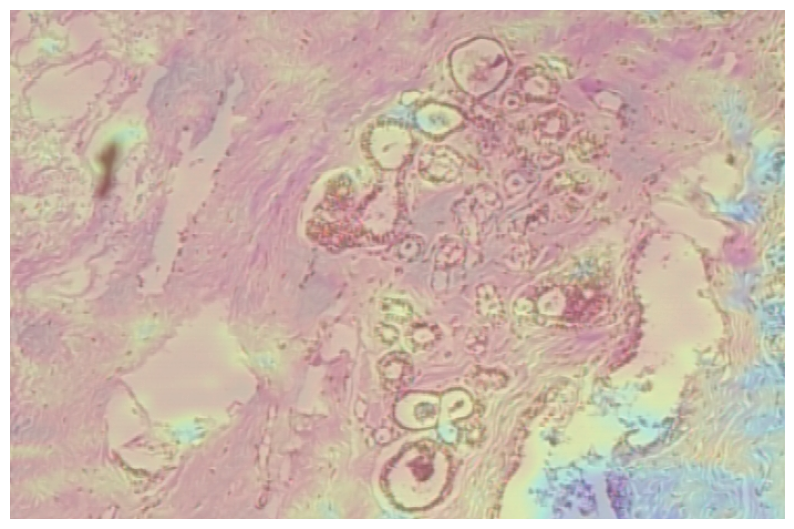}
\\40X-ScoreCAM
\end{minipage}
\hfill
\begin{minipage}{0.18\textwidth}
\centering
\includegraphics[width=\textwidth]{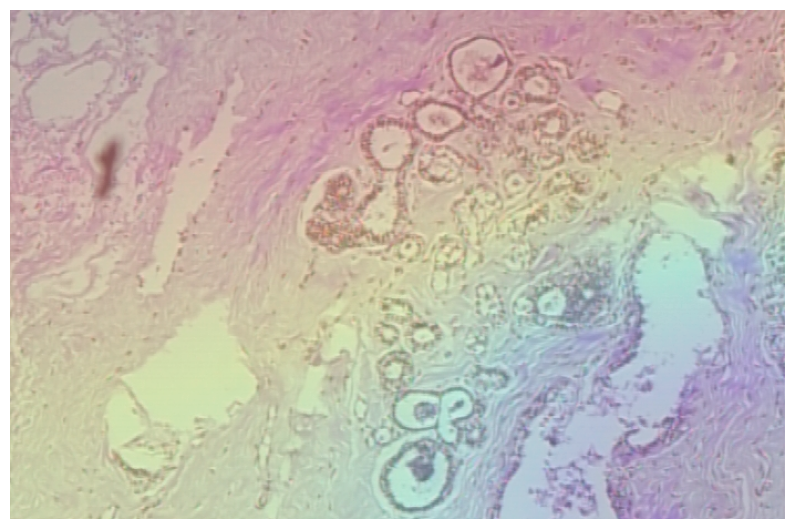}
\\40X-EligenCAM
\end{minipage}
\hfill
\begin{minipage}{0.18\textwidth}
\centering
\includegraphics[width=\textwidth]{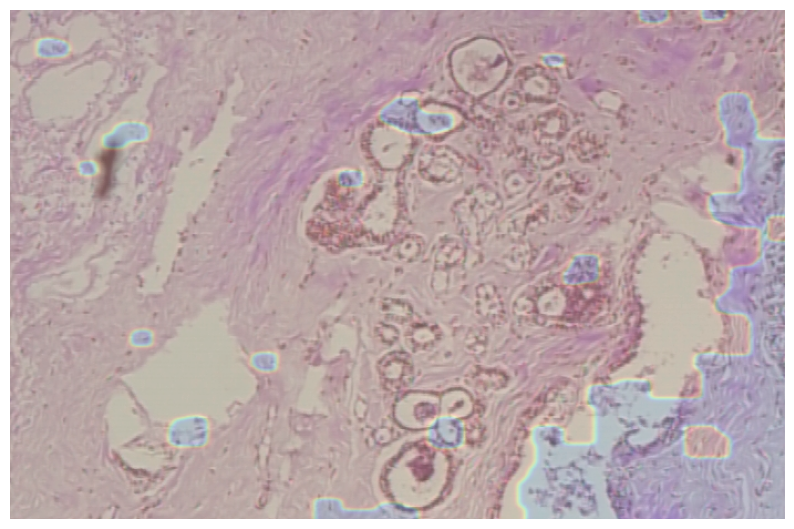}
\\40X-DT GradCAM
\end{minipage}

\vspace{0.5cm}

\begin{minipage}{0.18\textwidth}
\centering
\includegraphics[width=\textwidth]{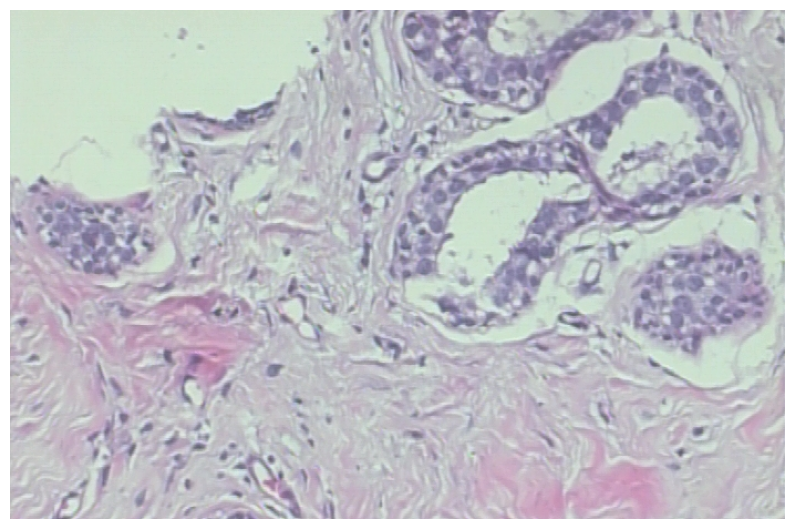}
\\100X-Original
\end{minipage}
\hfill
\begin{minipage}{0.18\textwidth}
\centering
\includegraphics[width=\textwidth]{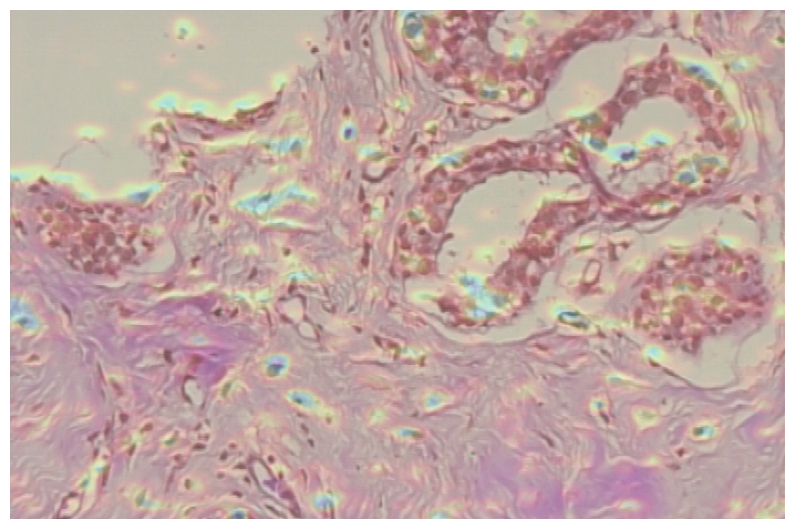}
\\100X-GradCAM
\end{minipage}
\hfill
\begin{minipage}{0.18\textwidth}
\centering
\includegraphics[width=\textwidth]{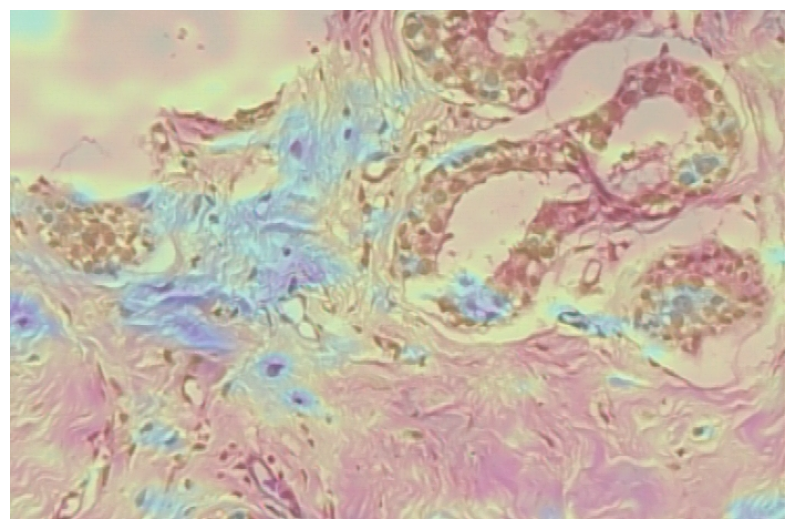}
\\100X-ScoreCAM
\end{minipage}
\hfill
\begin{minipage}{0.18\textwidth}
\centering
\includegraphics[width=\textwidth]{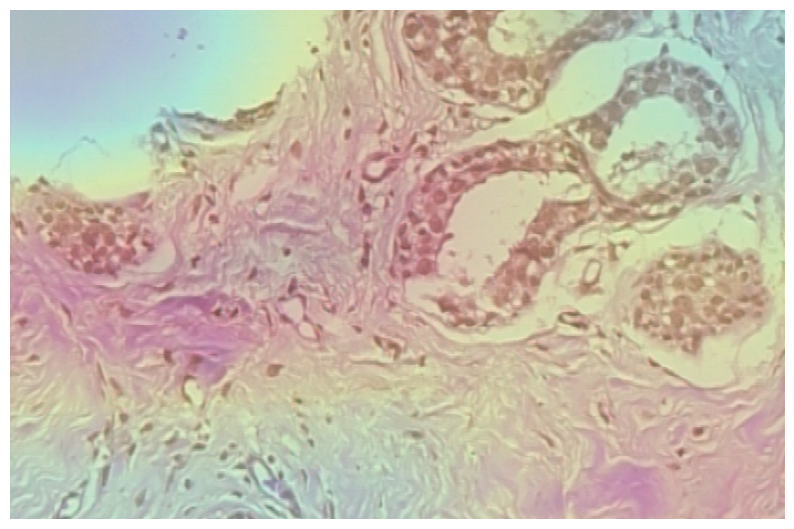}
\\100X-EligenCAM
\end{minipage}
\hfill
\begin{minipage}{0.18\textwidth}
\centering
\includegraphics[width=\textwidth]{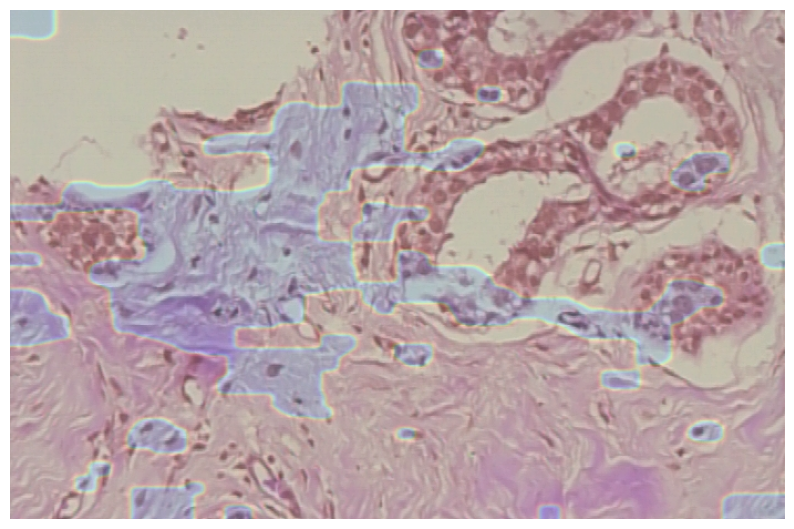}
\\100X-DT GradCAM
\end{minipage}

\begin{minipage}{0.18\textwidth}
\centering
\includegraphics[width=\textwidth]{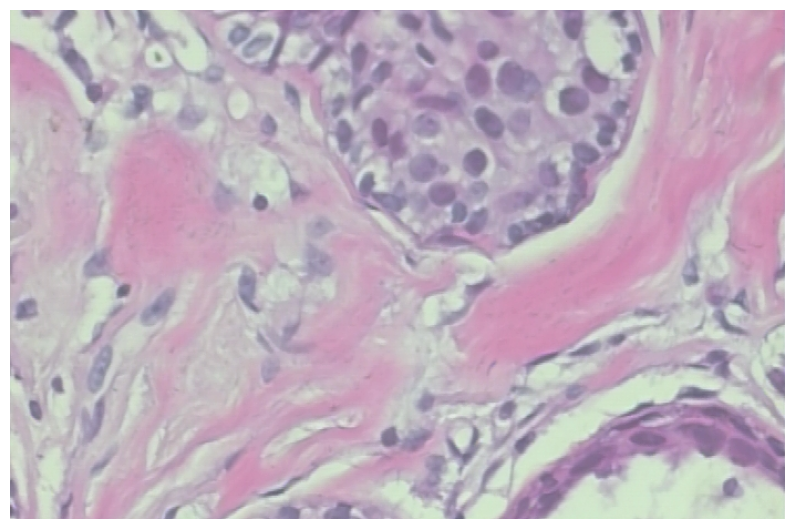}
\\200X-Original
\end{minipage}
\hfill
\begin{minipage}{0.18\textwidth}
\centering
\includegraphics[width=\textwidth]{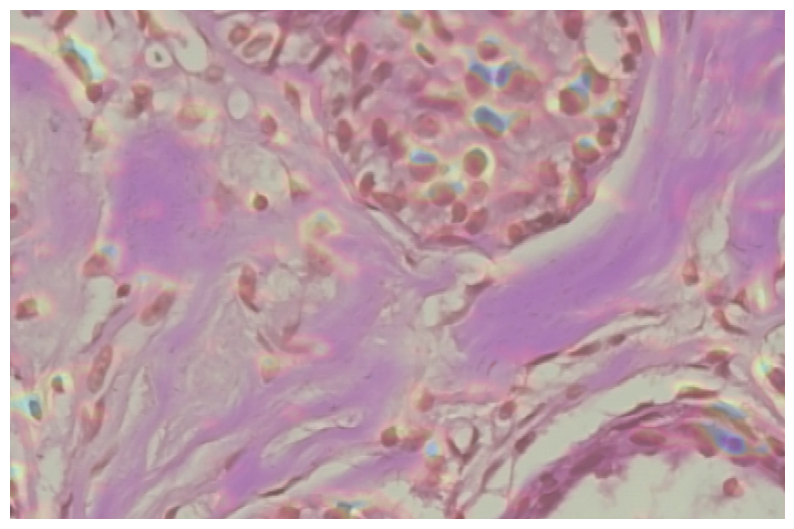}
\\200X-GradCAM
\end{minipage}
\hfill
\begin{minipage}{0.18\textwidth}
\centering
\includegraphics[width=\textwidth]{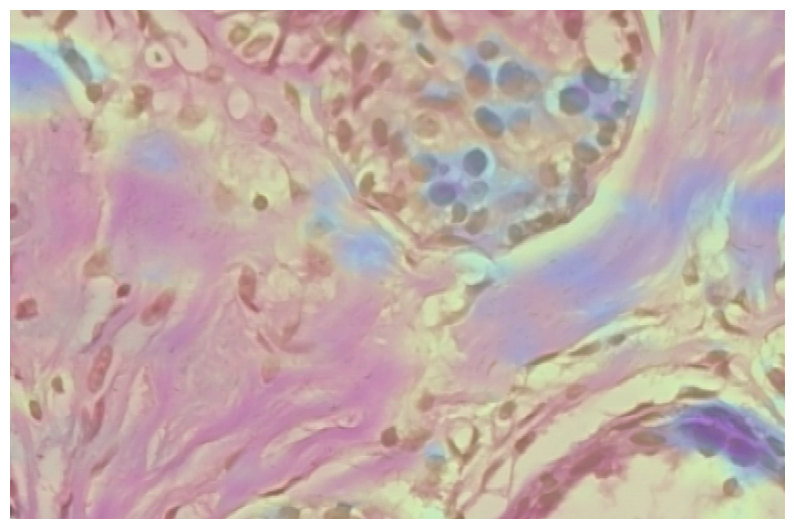}
\\200X-ScoreCAM
\end{minipage}
\hfill
\begin{minipage}{0.18\textwidth}
\centering
\includegraphics[width=\textwidth]{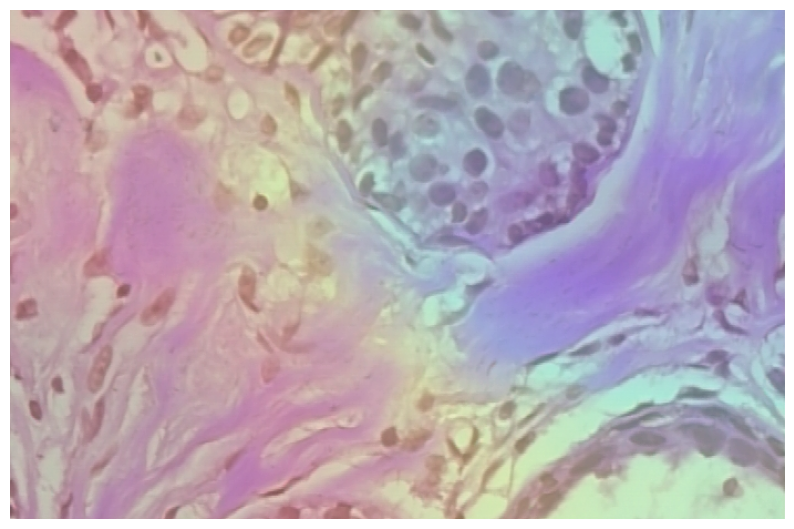}
\\200X-EligenCAM
\end{minipage}
\hfill
\begin{minipage}{0.18\textwidth}
\centering
\includegraphics[width=\textwidth]{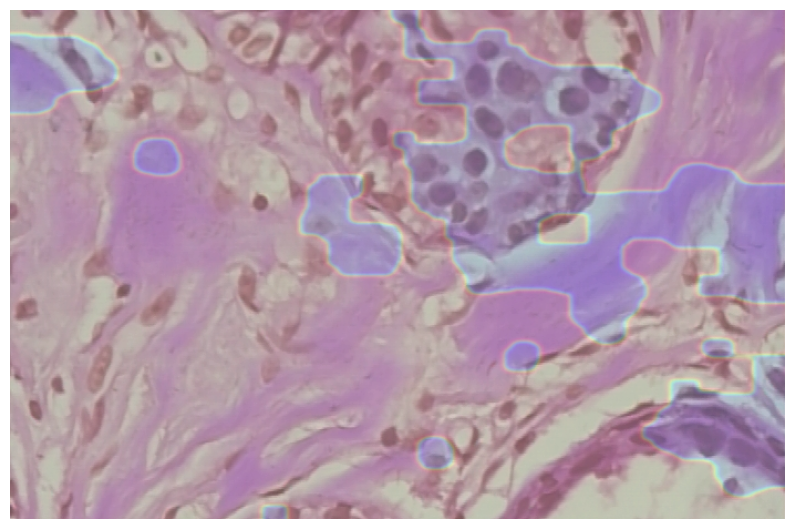}
\\200X-DT GradCAM
\end{minipage}

\begin{minipage}{0.18\textwidth}
\centering
\includegraphics[width=\textwidth]{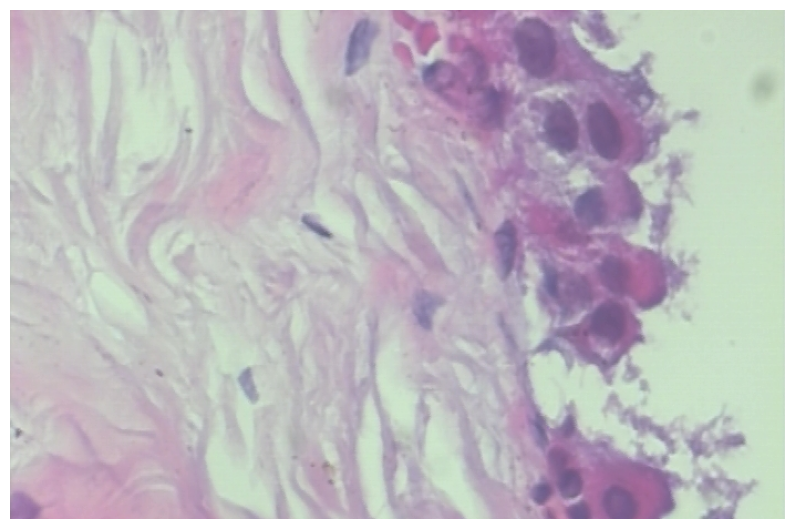}
\\400X-Original
\end{minipage}
\hfill
\begin{minipage}{0.18\textwidth}
\centering
\includegraphics[width=\textwidth]{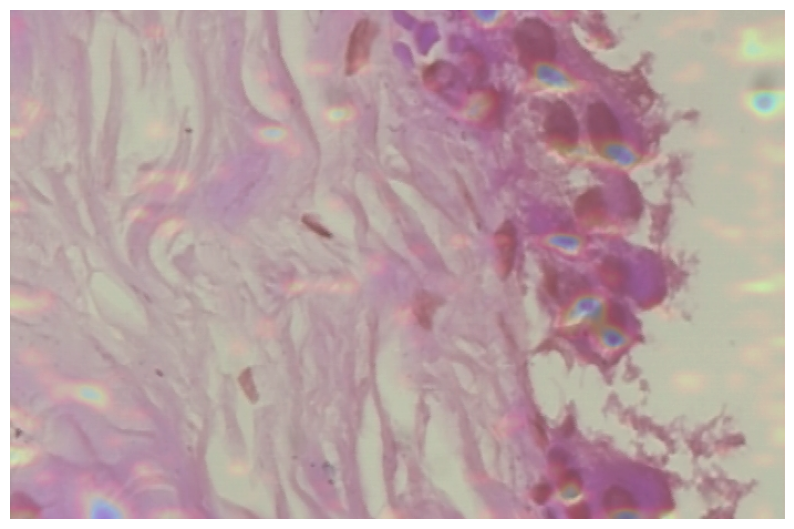}
\\400X-GradCAM
\end{minipage}
\hfill
\begin{minipage}{0.18\textwidth}
\centering
\includegraphics[width=\textwidth]{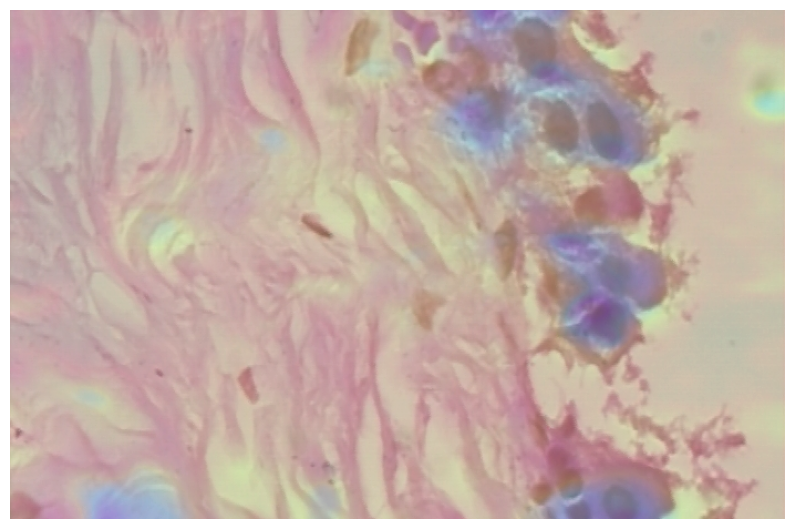}
\\400X-ScoreCAM
\end{minipage}
\hfill
\begin{minipage}{0.18\textwidth}
\centering
\includegraphics[width=\textwidth]{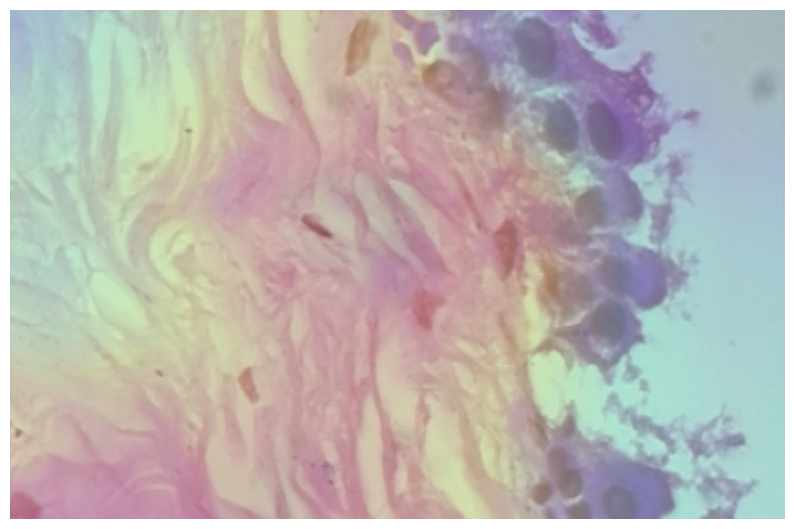}
\\400X-EligenCAM
\end{minipage}
\hfill
\begin{minipage}{0.18\textwidth}
\centering
\includegraphics[width=\textwidth]{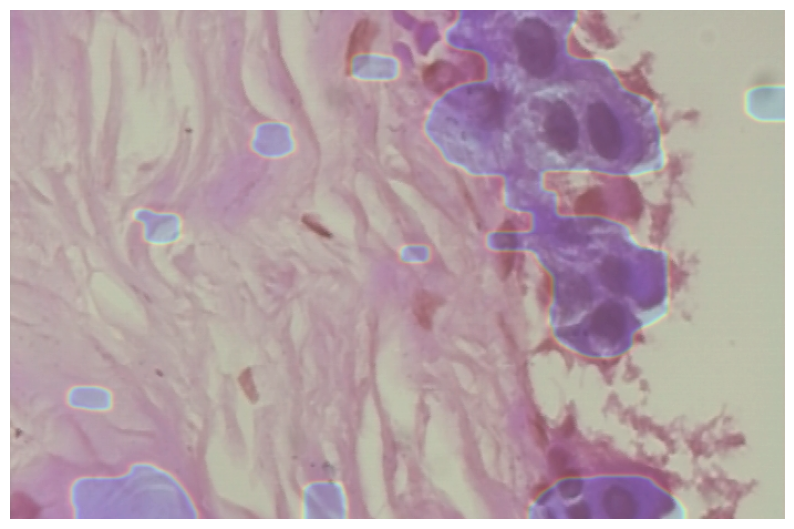}
\\400X-DT GradCAM
\end{minipage}

\caption{Comparison of DT Grad-CAM and Other Visualization Methods at Different Magnifications}
\end{figure*}

\subsection*{5. Thresholding and Morphological Operations}

The averaged heatmap is then binarized using the Otsu threshold:

\begin{equation}
L_{c}^{\text{Thresh}}(x, y) = 
\begin{cases} 
L_{c}^{\text{Avg}}(x, y) & \text{if } L_{c}^{\text{Avg}}(x, y) > T \\
0 & \text{otherwise}
\end{cases}
\end{equation}

After applying thresholding and morphological operations, the resulting DT Grad-CAM heatmap effectively highlights the most relevant regions with minimal noise and improved focus on critical areas. Through adaptive thresholding using Otsu's method and the weighted averaging of Grad-CAM maps generated from noisy inputs, DT Grad-CAM achieves clearer and more precise visualizations that isolate key regions. This enhanced focus on significant areas improves interpretability, making DT Grad-CAM especially valuable for applications that demand accurate localization and clear insights, such as breast cancer diagnosis in medical imaging.

After applying thresholding and morphological operations, the resulting DT Grad-CAM heatmap effectively highlights the most relevant regions with minimal noise and enhanced focus on critical areas. By leveraging adaptive thresholding through Otsu’s method and weighted averaging of Grad-CAM maps generated from noisy inputs, DT Grad-CAM delivers clearer and more precise visualizations that distinctly isolate key regions. This improved focus on significant areas enhances interpretability, making DT Grad-CAM an invaluable tool for applications requiring accurate localization and actionable insights, such as breast cancer diagnosis in medical imaging.

Table 13 summarizes the qualitative differences between DT Grad-CAM and Grad-CAM, emphasizing its advantages in adaptive thresholding, visualization clarity, interpretability, noise reduction, and diagnostic relevance.

\subsection{Comparison of DT Grad-CAM and Other CAM Methods}

The metrics IoU \cite{Abidin2024}, Dice \cite{Park2024}, Recall \cite{Kumar2024}, and F1 Score \cite{Cifci2023} provide a comprehensive assessment of the visualization quality and accuracy of DT Grad-CAM and Grad-CAM. IoU measures the overlap between predicted and true regions, indicating the accuracy of identifying relevant areas. The Dice Coefficient assesses the similarity between predicted and true regions, highlighting the method’s effectiveness in capturing key features. Recall evaluates the method's sensitivity in detecting relevant regions, while the F1 Score balances precision and recall, offering an overall measure of accuracy.

Based on Table 14, DT Grad-CAM consistently achieves higher scores across these metrics at various magnifications (40X, 100X, 200X, and 400X), demonstrating its superiority over Grad-CAM. This suggests that DT Grad-CAM outperforms Grad-CAM \cite{b60}, Score-CAM \cite{b61}, and Eigen-CAM \cite{Muhammad2021} in effectively highlighting critical regions across varying image scales. The improved scores demonstrate DT Grad-CAM’s effectiveness in enhancing visualization clarity and precision, particularly by emphasizing important areas while reducing irrelevant details.

Fig. 7 illustrates the visualizations provided by all four CAM methods—Grad-CAM, DT Grad-CAM, Score-CAM, and Eigen-CAM at different magnifications, highlighting DT Grad-CAM’s sharper and more focused results compared to the others. These enhancements make DT Grad-CAM particularly valuable for applications that require precise and interpretable visualizations, such as in medical imaging for breast cancer diagnosis.

\section{Summary and conclusions}

This paper introduces the DALAResNet50 model for classifying medical histopathology images, particularly addressing the challenges of limited-scale and imbalanced breast cancer datasets. By integrating a lightweight attention mechanism into the ResNet50 architecture, the model achieves superior performance, demonstrating robustness, better generalization, faster convergence, and effective prevention of overfitting.

Additionally, the proposed DT Grad-CAM method enhances interpretability by applying adaptive thresholding through Otsu’s method, resulting in clearer and more focused visualizations. This improves the model's transparency and aids medical experts in understanding critical features that drive classification decisions.

The combined use of DALAResNet50 and DT Grad-CAM ensures high classification accuracy and enhanced interpretability, which are essential for reliable breast cancer diagnosis. Future work will explore the application of this model in broader medical contexts, focusing on its integration with existing diagnostic workflows.

\begin{IEEEbiography}[{\includegraphics[width=1in,height=1.25in, clip,keepaspectratio]{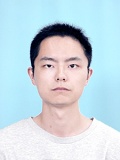}}]{SUXING LIU} received his Bachelor of Engineering from Anhui Agricultural University and his Master’s degree from Xi’an University of Technology, both in China. He is currently a university lecturer in China and is pursuing his Ph.D. at Mokwon University in South Korea. His research primarily focuses on medical image processing, and he has published several papers in the field of artificial intelligence.
\end{IEEEbiography}

\begin{IEEEbiography}[{\includegraphics[width=1in,height=1.25in,clip,keepaspectratio]{ 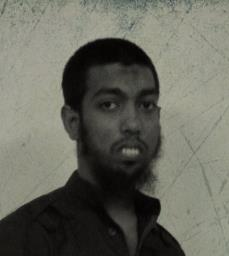}}]{Galib Muhammad Shahriar Himel} received his 1st BSc degree in Computer Science and Engineering from Ahsanullah University of Science and Technology (AUST) in the year 2016. Then he received his 1st MSc degree in Computer Science \& Engineering from United International University (UIU) in the year 2018. Then he received his 2nd BSc degree in Computing from the University of Greenwich (UoG), UK in 2021. After that, he received his 2nd MSc degree in Computer Science specializing in Intelligent Systems from American International University-Bangladesh (AIUB) in the year 2022. He completed his 3rd MSc in 
Applied Physics \& Electronics at Jahangirnagar (JU) in the year 2023. He has worked as a researcher both at Bangladesh University of Business and Technology (BUBT) and Independent University, Bangladesh 
(IUB). 

Currently, he is involved in several research studies related to Biomedical image processing using machine learning and is working as a PhD researcher at Universiti Sains Malaysia (USM). His research interests include Artificial Intelligence, Machine Learning, Bioinformatics, Bio-medical image analysis, and computer Vision. He has published several research articles on Artificial Intelligence, Machine Learning, and Biomedical Image Processing.

\end{IEEEbiography}

\begin{IEEEbiography}[{\includegraphics[width=1in,height=1.25in, clip,keepaspectratio]{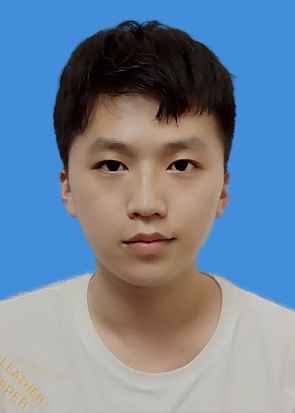}}]{JIAHAO WANG}  received his Bachelor's degree in Network Engineering from Fuzhou University(FZU) in 2018. He then received a master's degree in computer technology from Liaoning University of Technology(LUT)
 in 2022. He worked as an algorithm engineer at China's Qianxin Network security company. Currently, he is working on deep learn-related algorithms at Universiti Sains Malaysia (USM), where his research interests include artificial intelligence, deep learning, large language models, graph neural networks, contrast learning, and recommendation algorithms. He has published several research articles on artificial intelligence, machine learning, and recommendation algorithms.
\end{IEEEbiography}

\EOD

\end{document}